\documentclass[aps,twocolumn,showpacs,preprintnumbers,prb]{revtex4-1}
\usepackage{amsmath,amssymb}
\bibpunct{[}{]}{,}{n}{}{}
\usepackage{comment}
\usepackage{graphicx}
\usepackage{braket}
\usepackage{enumitem}
\usepackage{float}
\usepackage{color}
\def\v{\mathbf{v}}

\def\cL{\mathcal{L}}
\def\cH{\mathcal{H}}

\def\cO{\mathcal{O}}

\def\cV{\mathcal{V}}

\def\cS{\mathcal{S}}

\def\cU{\mathcal{U}}

\def\>{\rangle}
\def\<{\langle}

\newcommand\+{\dagger}
\newcommand\x{\mathbf{x}}

\newcommand\mk{\mathbf{k}}
\newcommand\mx{\mathbf{x}}
\newcommand\my{\mathbf{y}}
\newcommand\mq{\mathbf{q}}

\newcommand*\diff{\mathop{}\!\mathrm{d}}
\renewcommand{\vec}[1]{\mathbf{#1}}

\newcommand{\be}{\begin{equation}}
\newcommand{\ee}{\end{equation}}
\newcommand{\bea}{\begin{eqnarray}}
\newcommand{\eea}{\end{eqnarray}}
\newcommand{\p}{\partial}

\newcommand{\cE} {\mathcal{E}}

\usepackage[colorlinks,bookmarks=true,citecolor=blue,linkcolor=red,urlcolor=blue]{hyperref}
\usepackage{hyperref}

\begin{document}
	
	\title{Quantum Boltzmann equation for bilayer graphene}

	\author{Dung X. Nguyen}
	\affiliation{Rudolf Peierls Centre for Theoretical Physics, Parks Road, Oxford, OX1 3PU, UK}
	\author{Glenn Wagner}
	\affiliation{Rudolf Peierls Centre for Theoretical Physics, Parks Road, Oxford, OX1 3PU, UK}
	\author{Steven H. Simon}
	\affiliation{Rudolf Peierls Centre for Theoretical Physics, Parks Road, Oxford, OX1 3PU, UK}
	
	\begin{abstract}
	A-B stacked bilayer graphene has massive electron and hole-like excitations with zero gap in the nearest-neighbor hopping approximation.  In equilibrium, the quasiparticle occupation approximately follows the usual Fermi-Dirac distribution.    In this paper we consider perturbing this equilibrium distribution so as to determine DC transport coefficients near charge neutrality.   We consider the regime $\beta |\mu| \lesssim 1$ (with $\beta$ the inverse temperature and $\mu$ the chemical potential) where there is not a well formed Fermi surface. Starting from the Kadanoff-Baym equations, we obtain the quantum Boltzmann equation of the electron and hole distribution functions when the system is weakly perturbed out of equilibrium. The effect of phonons, disorder, and boundary scattering for finite sized systems are incorporated through a generalized collision integral. The transport coefficients, including the electrical and thermal conductivity,  thermopower, and shear viscosity, are calculated in the linear response regime. We also extend the formalism to include an external magnetic field. We present results from numerical solutions of the quantum Boltzmann equation. Finally, we derive a simplified two-fluid hydrodynamic model appropriate for this system, which reproduces the salient results of the full numerical calculations.

	\end{abstract}
	
	\maketitle
	
	\section{Introduction}
	Graphene is a remarkable material that has generated enormous interest in both the theoretical and experimental community since its discovery in 2004 \cite{Geim2009}.   While there are many reasons for interest in this unusual material, one particularly exciting feature is that it is possible to produce graphene samples with extreme purity and thus observe an electronic regime that had not previously been explored ---  the hydrodynamic regime \cite{lucas2018hydrodynamics,Ho2018,Bandurin1055,Levitov2016,Mendoza2013}. In this regime the shortest scattering time is that for electron-electron collisions, and  collisions with phonons and impurities are subdominant.   As a result, semi-classical kinetic theory reduces to a form of quantum hydrodynamics \cite{Irving1951}.   
	
    In general, quantum Hydrodynamics	(QH) describes the dynamics of systems that vary slowly in space and time. The foundation of QH is the ability to obtain a set of conservation laws for the electron liquid. These conservation laws are derived from the quantum Boltzmann equation (QBE), which is the equation of motion for the electron fluid's phase space distribution function \cite{Erdos2004,Kadanoff}.  
    The QH approach has been extremely successful in studying electron plasmas,  fractional quantum Hall fluids, \cite{Passot2005,Wiegmann2013,Son2013}, and now also the electron fluid of graphene \cite{Fritz2008,lucas2018hydrodynamics}.   In the current work we will derive QH equations from the QBE for the case of bilayer graphene (BLG) near charge neutrality (CN) where there is no well defined Fermi surface.   We note, however, that our kinetic theory formalism is more general than QH.

    Whereas QBE and QH have been studied extensively in monolayer graphene \cite{Fritz2008,Mueller2008,Mueller2008b,Lux2012}, they have not been as well studied in the case of bilayer graphene.    There are, however, several reasons why the BLG case should be of substantial interest.    Firstly, the band structure of BLG is fundamentally different from that of monolayer graphene.   	In the nearest-neighbor hopping approximation, A-B stacked bilayer graphene has quadratic bands of electron and hole-like excitations at low energies \cite{Neto:2007} which touch at zero energy.  This interesting band structure, which was confirmed experimentally in Ref \cite{Nam2017}, provides the unique quantum transport properties of BLG \cite{McCann2013}. 
	
	A second reason that BLG is now of interest is due to recent experimental advances that have allowed measurements with unprecedented precision ---  in particular \cite{Nam2017} reports measurements of the electrical conductivity of BLG. These advances have been possible due to the development of suspended BLG devices \cite{Nam2017,sBLG1,sBLG2,sBLG3,sBLG4,sBLG5,sBLG6,sBLG7,sBLG8}. As with monolayer graphene, BLG on a substrate suffers an inhomogeneous potential, which can lead to charge-puddle physics and superlattice effects. Suspended samples, in comparison, are far cleaner.   The current limitation on suspended samples is on the size of these devices, however recent BLG devices have achieved sizes longer than the disorder scattering length \cite{Ki2013}.
	
	Further, due to the low impurity scattering rates in clean samples, there has also been recent interest in studying the viscosity of the electron fluid in materials such as graphene \cite{Levitov2016,Torre2015}. Some signatures of electron viscosity, such as negative non-local resistance, have already been measured experimentally in monolayer graphene\cite{Levitov2016a}.  The extension of these experiments to BLG seems natural. 
	
	Finally, measurements of the thermal conductivity of suspended single-layer graphene have been performed \cite{Baladin2008} and it may be possible to extend this study to BLG as well. 
	
	In this paper, we develop the QBE formalism for calculating transport properties of BLG which can in principle be compared against existing and future transport experiments. The analogous formalism for the electrical conductivity of single-layer graphene was worked out in \cite{Fritz2008}. This work was later extended to study Coulomb drag between two monolayers of graphene \cite{Lux2012} and BLG exactly at charge neutrality (CN) \cite{Lux2013}. We extend this work away from CN. To do so, we start with the derivation of the quantum transport operators including the charge current, the heat current and the stress tensor in terms of electron and hole fields. We include the Zitterbewegung contribution to these operators. Taking the expectation value of these operators in the DC limit, we obtain the main results for the electrical current \eqref{eq:currentexpect},  heat current \eqref{eq:Qcurrentexpect} and stress tensor \eqref{eq:Tijexpect}. 
	We then calculate the collision integral, which in general will have four contributions. A central result of our work is the collision integral \eqref{eq:CoItot}, which contains contributions from the Coulomb interactions \eqref{eq:CoI}, impurity scattering \eqref{eq:CoD}, scattering off the boundary \eqref{eq:CoS} and phonon scattering \eqref{eq:I_phonon}.
	
	Once we have derived the QBE formalism, we can study the linear response of the system to perturbations. In particular, we study the behavior under an applied external electric field, thermal gradient, and straining motion in order to calculate the electrical conductivity, thermal conductivity, and viscosity respectively. Finally, in anticipation of 
    future experiments, we also  consider the behavior of the transport properties in an applied magnetic field which leads to, e.g., nonzero Hall-conductivity.    The inclusion of a magnetic field is possible for both electrical conductivity and thermal conductivity calculations. 
	
	Since the dominant scattering mechanism is electron-electron, and hole-hole, we should be able to represent the transport with a two-fluid hydrodynamics --- where the electron fluid and the hole fluid are individually in a thermal equilibrium, each having a well defined temperature, chemical potential, and velocity.   As such we use the QBE to derive a two-fluid model which describes the evolution of the mean fluid velocities of the electron and hole fluids on timescales long compared to the electron-electron collision time. These are given by equations \eqref{eq:fluid1} and \eqref{eq:fluid2}. The two-fluid model includes Coulomb drag between the electron and hole fluids and the momentum-relaxing scattering from scattering with phonons. This then allows us to derive simple analytical expressions for the transport properties.

	The plan of the paper is as follows. In section \ref{sec:wf} we review the electronic structure of bilayer graphene and the low-energy effective Hamiltonian. Section \ref{sec:Coul} deals with the Coulomb interaction between electrons, in particular the screening thereof in the RPA approximation. We perform the calculation in flat space first and then generalize to curved space so that we can later calculate the stress tensor, which gives the response to a change in the metric. In section \ref{sec:Conserved_currents} we calculate the conserved currents. We then derive the kinetic equation in section \ref{sec:Kinetic_Equation}. Section \ref{sec:Bfield} explores the effect of a constant magnetic field on the kinetic equation. In \ref{sec:Coll} we write down the collision integral. Then in section \ref{sec:sym} we discuss the symmetries of the collision integral. We introduce the two-fluid model in \ref{sec:2fm} and derive analytical expressions for the transport properties in terms of the fundamental parameters of the problem. In section \ref{sec:num_res} we evaluate the collision integral numerically and show the results for some transport properties of interest. We compare to the results from the two-fluid model and find good agreement. Detailed calculations are left for the Appendices.

	\section{Review of the electronic structure of bilayer graphene}
	To start off, we  briefly review the derivation of the band structure and the explicit expression of the wave function in BLG. This part serves to make this work self-contained and to introduce notation. 
	
	\label{sec:wf}
	\subsection{Hamiltonian}
	
	\begin{figure}
		\centering
		\includegraphics[width=0.4\textwidth]{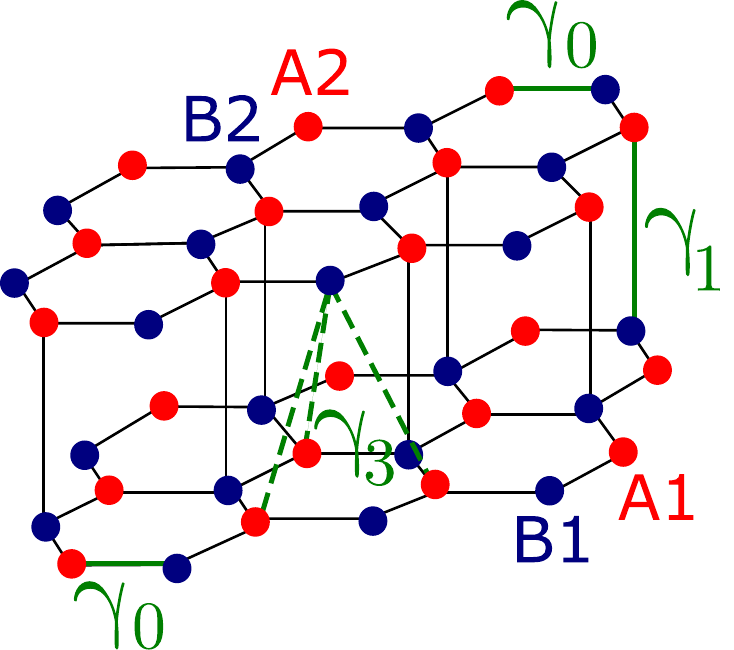}
		\caption{Sketch of the bilayer graphene lattice used for the tight-binding Hamiltonian. There are two layers 1 and 2 and in each layer there are two inequivalent sites per unit cell labelled A and B. The couplings $\gamma_0$, $\gamma_1$ and $\gamma_3$ are defined in the figure.}
		\label{fig:tight-binding}
	\end{figure}
	
	The tight-binding Hamiltonian of A-B stacked bilayer graphene with the coupling defined in Fig. \ref{fig:tight-binding} and external gauge field $A_\mu$ has the explicit form \cite{McCann2006,Misumi2008,McCann2013}
	\begin{align}
	\label{eq:Ham}
	H_\xi=\xi\begin{pmatrix}
	0 &v_3 \pi &0 &v_F \pi^\dagger\\v_3 \pi^\dagger &0 &v_F \pi &0\\0 &v_F \pi^\dagger &0&\xi \gamma_1\\v_F \pi &0 &\xi \gamma_1 &0
	\end{pmatrix}-eA_0 I,
	\end{align}
	where 
	the velocity $v_3$ is given by $v_3=\frac{\sqrt 3}{2}a\gamma_3 /\hbar$, where $a$ is the lattice constant, 	and the Fermi-velocity is given by $v_F=\frac{\sqrt 3}{2}a\gamma_0 /\hbar$ \footnote{We omit the spin indices here and will take the spin degeneracy into account later on, when we introduce the fermion flavor. The Zeeman splitting is negligible for the fields that we consider here ($B<100$Gauss).}.  Here, 
  $\xi=1$ corresponds to $K$ valley with corresponding wave function \footnote{Here we use the integral over $\mathbf{k}$ as the summation over $\mathbf{k}$ about each Dirac cone $K$ or $K'$ with a cut off $\Lambda$ of order $2\pi/a$. The physical results in our paper do not depend on the cut off since we consider the physics in the long wavelength limit where $k\ll \Lambda$.}
	\begin{align}
	\psi_K(\mathbf{x})&=\begin{pmatrix}
	\varphi_{A_1}(\mx)\\\varphi_{B_2}(\mx)\\\varphi_{A_2}(\mx)\\\varphi_{B_1}(\mx)
	\end{pmatrix}=\int \frac{d^2k}{(2\pi)^2}\psi_K(\mathbf{k})e^{i\mk\mx},
	\end{align}
	and $\xi=-1$ corresponds to the $K'$ valley with corresponding wave function $
	\psi_{K'}(\mathbf{x})=(
	\varphi_{B_2}(\mx),\varphi_{A_1}(\mx),\varphi_{B_1}(\mx),\varphi_{A_2}(\mx))$.
	We have defined the momentum operator and its holomorphic and anti-holomorphic notation
	\begin{align}
	p_i=-i\hbar \partial_i -eA_i\\
	\pi=p_x +i p_y, \qquad \pi^\dagger=p_x -i p_y.
	\end{align}
	$e<0$ is the electron charge. In the following, we will set $v_3=0$, since we are only interested in the quadratic bands (see Ref.~\cite{McCann2013} for details).

	\subsection{Effective Hamiltonian}
	Since the Hamiltonian \eqref{eq:Ham} provides information about both high energy and low energy states, it will be useful to create a low-energy effective Hamiltonian. 
	To simplify our model, we consider only the low-energy bands near the $K$ valley. In the long wavelength limit $v_F k \ll \gamma_1$, one can derive for $\xi=1$
	\begin{equation}
	\label{eq:Hameff1}
	H_{K}=-\frac{1}{2m}\begin{pmatrix} 
	0  &(\pi^\dagger)^2 \\
	\pi^2 & 0 
	\end{pmatrix}, \qquad \psi_K=\begin{pmatrix} 
	\varphi_{A_1}(\mathbf{x})   \\
	\varphi_{B_2}(\mathbf{x})
	\end{pmatrix},
	\end{equation}
	where $m=\frac{\gamma_1}{2v_F^2}$. 
	The wave function is given by
	\begin{equation}
	\label{eq:wfeK}
	\psi^\lambda_K=\frac{1}{\sqrt{2}}\begin{pmatrix} 
	-\lambda e^{-2i\theta_{\mathbf{k}}}   \\
	1
	\end{pmatrix}, \qquad \epsilon^\lambda_{\mathbf{k}}=\lambda\frac{k^2}{2m},
	\end{equation}
	where $\theta_{\mk}$ is the angle between the vector $\mk$ and the $x$-axis. $\lambda=-1$ denotes electrons in the valence band and $\lambda=1$ denotes electrons in the conduction band. In the low energy limit, we only consider the electrons appearing at sites $A_1$ and $B_2$.
	In the following sections, we will omit the spin indices for simplicity and consider them back in the counting factors. 
	Similarly, we can derive the effective Hamiltonian near the $K'$ valley, the only difference is that the wave function is now $\psi_{K'}=(	\varphi_{B_2}(\mathbf{x}),	\varphi_{A_1}(\mathbf{x}))$
	
	\section{Coulomb interaction and screening}
	\label{sec:Coul}
	Coulomb screening is the damping of the electric field due to mobile charge carriers which are quasi-particles and quasi-holes. As a result of screening, the long-range Coulomb interaction becomes short-range. In this section, we will calculate the screening effect of the Coulomb interactions in BLG, or in other words we will calculate the screening momentum $q_{TF}$.  
	
	\subsection{Charge density operator}
	The Hamiltonian \eqref{eq:Ham} shows explicitly the coupling of BLG with an external gauge field. The free Lagrangian density is given by 
	\begin{equation}
	\label{eq:freeL}
	\cL_\xi=i\hat \Psi^\dagger_{\xi}\overleftrightarrow{\partial_t} \hat \Psi_\xi -\hat\Psi^\dagger_{\xi} H_\xi \hat \Psi_\xi,
	\end{equation}
	where $\overleftrightarrow{\partial_t}=\frac{1}{2}(\overrightarrow{\partial_t}-\overleftarrow{\partial_t})$. The field operator in the $K$ valley in second quantization language is given by 
	\begin{equation}
	\hat{\Psi}_K(\mx)=\int \frac{d^2k}{(2\pi)^2}\begin{pmatrix}
	c_{K;A_1}(\mk)\\c_{K;B_2}(\mk)\\c_{K;A_2}(\mk)\\c_{K;B_1}(\mk)
	\end{pmatrix}e^{i\mk\mx},
	\end{equation}
	where the operator $c_{K;a}(\mk)$ is the annihilation operator of an electron on the sublattice $a$ at momentum $\mathbf{K+k}$. Similarly, the field operator in the $K'$ valley in second quantization language can be derived. 	The free action in flat space is given by
	\begin{equation}
	\cS_{\xi}=\int d^3 x \cL_\xi.
	\end{equation}
	The total number density operator in both valleys is given by the definition 
	\begin{equation}
	\rho(\mathbf{x})=\sum_{\xi}\rho_{\xi}(\mathbf{x})=\sum_{\xi,s}\frac{1}{e}\textbf{}\frac{\delta \cS_{\xi}}{\delta A_0}(\mx)=\sum_{\xi,s}\hat{\Psi}^\dagger_\xi(\mx)\hat{\Psi}_{\xi}(\mx).
	\end{equation}
	From the wave function of the electron and hole bands at low energies, we can derive the transformation of the field operator
	\begin{equation}
	c_{\pm K}(\mk)=\frac{1}{\sqrt 2} \left(\mp e^{-2i\theta_{\mk}}c_{K;A_1}(\mk)+c_{K;B_2}(\mk)\right),
	\end{equation}
	and similarly for the $K'$ bands.

	Combining the spin index and the valley index to a flavor index, we obtain the effective low-energy density
	\begin{multline}
	\rho^{\text{eff}}(\mq)=\sum_{f}\int \frac{d^2 k}{(2\pi)^2}\frac{1}{2}\\\times\sum_{\lambda,\lambda'}c^\dagger_{\lambda f}(\mk-\mq)c_{\lambda' f}(\mk)\left(1+\lambda \lambda' e^{-2 i(\theta_{\mk-\mq}-\theta_{\mk})}\right).
	\label{eq:rho_eff}
	\end{multline}
	We can separate the effective charge density \eqref{eq:rho_eff} into a normal part where $\lambda=\lambda'$ and the \textit{Zitterbewegung} part where $\lambda=-\lambda'$. The homogeneous contribution to the charge density $\rho^{\text{eff}}(\mathbf{0})$ is only due to the normal part. The effective Coulomb interaction of the effective theory is given by 
	\begin{equation}
	\label{eq:Coulomb}
	\hat{V}_C=\frac{1}{2}\int \frac{d^2 q}{(2\pi)^2} V_C(\mq)\rho^{\text{eff}}(-\mq)\rho^{\text{eff}}(\mq).
	\end{equation}
	The screened Coulomb interaction $V_C(q)$ will be calculated in the next subsection. The bare Coulomb interaction is
	\begin{equation}
	V(q)=\frac{2\pi\alpha}{q}
	\label{eq:bare}
	\end{equation}
	and $\alpha=\frac{e^2}{4\pi\varepsilon_0}$.
	
	\subsection{Screening in flat background metric}
	We need to account for the screening of the long-range Coulomb interaction by the mobile charge carriers. In the random phase approximation (RPA) the dressed interaction is given by 
	\begin{equation}
	V_C(q,\omega)=\frac{V(q,\omega)}{1-\varPi^0(q,\omega) V(q,\omega)},
	\label{eq:RPA}
	\end{equation}
	where $\varPi^0(q,\omega)$ is the bare susceptibility. In order to calculate $\varPi^0(q,\omega)$, we need to calculate the fermion loop of BLG at a finite temperature and at a given chemical potential $\mu$. This is the textbook Lindhart calculation and in the regime $\beta\mu\lesssim1$ and $\beta q^2/m\lesssim 1$ we use the approximate result from Ref \cite{Lv2010}
	\begin{equation}
	\varPi^0(\mq,0)\approx-\frac{m N_f}{2\pi}(1+\frac{\beta q^2}{12m}).
	\end{equation}
	This equation starts deviating from the full result in Ref \cite{Lv2010} for $\beta q^2/m>1$, however large momentum transfer is suppressed by the Fermi occupation factor. We have that the screened potential is given according to \eqref{eq:RPA} by 
	\begin{equation}
	V_C(q)=\frac{2\pi \alpha}{q+q_{TF}(q)}
	\label{eq:screened_V}
	\end{equation}
	with the screening momentum 
	\begin{equation}
	\label{eq:qTF}
	q_{TF}(q)=-\Pi^0(\mq,0) 2\pi \alpha.
	\end{equation}
	For $\beta\mu\lesssim 1$, the typical momentum is $k_T=\sqrt{2 k_B T m/\hbar^2}\ll q_{TF}$ for any realistic temperature, so we can safely approximate
	\begin{equation}
	V_C(q)=\frac{2\pi \alpha}{q_{TF}(q)}.
	\end{equation}
	
	\subsection{Screening in a homogeneous metric}
	In order to calculate the stress tensor, we need to generalize the formalism to curved space. For a homogeneous metric $g_{ij}=\delta_{ij}+\delta g_{ij}$, we can follow the steps in the last subsection and obtain the screened Coulomb interaction as 
	\begin{align}
	\label{eq:CoulombCur}
	\tilde{\cV}(\mq,0)=\frac{2\pi \alpha}{\sqrt{g}}\frac{1}{|\mq|+q_{TF}(|\mq|)},
	\end{align}
	where $|\mq|=\sqrt{g^{ij}q_iq_j}$ and $g=\textrm{det}(g_{ij})$ and where $q_{TF}(|\mq|)$ takes the form of \eqref{eq:qTF}. The detailed derivation of equation \eqref{eq:CoulombCur} is given in Appendix \ref{sec:CoulbCur}. Equation \eqref{eq:CoulombCur} is a new result of this paper and is required in order to calculate the stress tensor operator in the next section.

	\section{Conserved current operators}
	\label{sec:Conserved_currents}
	In order to calculate the transport coefficients, we need to start by deriving the conserved current operators in the effective theory in second quantization language. The detailed derivations for the energy current and the stress tensor operators of BLG are new contributions of this paper. As has been shown in Refs \cite{Fritz2008} and \cite{Lux2013}, there are \textit{Zitterbewegung} contributions to the charge current operator of graphene as well as BLG. To obtain the DC transport coefficients of BLG, one can neglect the Zitterbewegung part which is just the contribution from the off-diagonal component of the Green's function in the generalized Kadanoff-Baym formalism \footnote{The detailed discussions of the Kadanoff-Baym formalism are left for Appendix \ref{sec:BoltzA}.}. However, this contribution will be necessary for studying the quantum transport at finite frequency and momentum using the QBE and we therefore include it for future extensions of this work. In this paper, we are only interested in spatially-independent current operators. 
	
	\subsection{Charge current operator}
	The current operator is by definition 
	\begin{equation}
	\label{eq:chargecurrent}
	J^i_{\xi}(\mx)=\frac{\delta \cS_\xi}{\delta A_i (\mx)}=-v_Fe\sum_{s} \xi\hat{\Psi}^\dagger_{\xi}(\mx)\begin{pmatrix}
	0 & (\sigma^i)^{\dagger} \\ 
	(\sigma^i)^{\dagger}& 0
	\end{pmatrix}\hat{\Psi}_{\xi}(\mx)
	\end{equation}
	where $s$ stands for spin.
	The current density is given by 
	\begin{equation}
	\mathbf{J}(\mx)=\sum_{\xi}\mathbf{J}^{\text{I}}_{\xi}(\mx)+\mathbf{J}^{\text{II}}_{\xi}(\mx)
	\end{equation}
	where $\mathbf{J}^{\text{I}}_{\xi}(\mx)$ is the contribution of quasi-particle and quasi-hole flow, and  the operator $\mathbf{J}^{\text{II}}_{\xi}(\mx)$ creates a quasi-particle-quasi-hole pair. 
	Using the explicit wave function of low-energy modes \eqref{eq:WaveK1e}, \eqref{eq:WaveK2e} and equation \eqref{eq:chargecurrent}, one can derive the spatially independent current operator 
	\begin{equation}
	\label{eq:chargecurrent1}
	\mathbf{J}^{\text{I}}(\mq=0)=\frac{e}{m}\sum_{f}\sum_{\lambda}\int \frac{d^2k}{(2\pi)^2} \lambda \mk c^\dagger_{\lambda f}(\mk) c_{\lambda f}(\mk),
	\end{equation}
	where we combined the spin index $s$ and valley index $\xi$ to flavor index $f$. Similarly, the \textit{Zitterbewegung} current operator is given by 
	\begin{multline}
	\mathbf{J}^{\text{II}}(\mq=0)=-i\frac{e}{m}\sum_{f}\int \frac{d^2k}{(2\pi)^2}(\hat{z}\times \mk) \\\times\left[c^\dagger_{+ f}(\mk) c_{- f}(\mk)-c^\dagger_{- f}(\mk) c_{+ f}(\mk)\right].
	\end{multline}
	\subsection{Energy current operator}
	The heat current is related to the energy current via 
	\begin{equation}
	\label{eq:heat}
	\mathbf{J}^{Q}(\mq=0)=\mathbf{J}^E(\mq=0)-\frac{\mu}{e} \mathbf{J}(\mq=0),
	\end{equation}
	where $\mu$ is the chemical potential and so we now calculate the energy current, which has contributions from both the kinetic and interaction energy terms in the Hamiltonian. We will follow Ref \cite{Jonson1980} in deriving the energy current operator. The conservation of energy gives us the continuity equation 
	\begin{equation}
	\label{eq:Encont}
	    \mathbf{\partial}\cdot \mathbf{J}^{E}(\mx,t)+ \dot{\cE}(\mx,t)=0,
	\end{equation}
	where $\mathbf{J}^{E}$ is the total energy current, which includes both kinetic and interaction contributions, $\cE$ is the energy density. We will use equation \eqref{eq:Encont} as the definition of the energy current.
	\subsubsection{Kinetic contribution}
	
	The kinetic energy density operator is given by 
	\begin{equation}
	\cE_{kin}(\mx)=\sum_{\xi}\mathbf{H}_{\xi}(\mx)=\sum_{\xi}\hat{\Psi}^\dagger_\xi(\mx)\overleftrightarrow{H}_{\xi}\hat{\Psi}_\xi(\mx)
	\end{equation} 
	where $\overleftrightarrow{H}_{\xi}$ means we replace $\partial_i$ in $H_{\xi}$ by $\overleftrightarrow{\partial}_i=\frac{1}{2}(\overrightarrow{\partial}_i-\overleftarrow{\partial}_i)$. We can write down the kinetic energy density in momentum space by Fourier transformation
	\begin{align}
	\mathbf{H}_{\xi}(\mq)&=\int d^2 \mx e^{-i\mq\mx}h_{\xi}(\mx)\\&=\int \frac{d^2 k}{(2\pi)^2}\hat{\Psi}^\dagger_\xi(\mathbf{k-q})\hat{\mathbf{H}}_{\xi}(\mk,\mq)\hat{\Psi}_\xi(\mk),\nonumber
	\end{align}
	where $\hat{\mathbf{H}}_{\xi}(\mk,\mq)$ is given explicitly as follows
	\begin{multline}
	\label{eq:hkin}
	\hat{\mathbf{H}}_{\xi}(\mk,\mq)=\\\xi\begin{pmatrix}
	0 &0 &0 &v_F (\bar{k}-\frac{1}{2} \bar q)\\0 &0 &v_F (k-\frac{1}{2}q) &0\\0 &v_F (\bar k -\frac{1}{2}\bar q) &0&\xi \gamma_1\\v_F (k-\frac{1}{2}q) &0 &\xi \gamma_1 &0
	\end{pmatrix},
	\end{multline}
	where we defined the holomorphic and anti-holomorphic vectors $X=X_1+i X_2,\qquad \bar{X}=X_1-iX_2$. Using Heisenberg's equation $\dot{\cE}_{kin}=-i[\dot{\cE}_{kin},\cH]$
	and the continuity equation of energy \eqref{eq:Encont} in momentum space we obtain the formula to determine the kinetic contribution to the energy current 
	\begin{equation}
	\label{eq:Energycurrent}
	\mq\cdot \mathbf{J}_{kin}^E(\mq)=[\cE_{kin}(\mq),\cH], 
	\end{equation}
	where the total Hamiltonian $\cH$ is defined in equation \eqref{eq:htot}. We leave the detailed calculation, which is quite technical, to Appendix \ref{sec:operaotsApp}. We only quote here the results after taking the limit $\mathbf{q}\rightarrow 0$ 
	\begin{equation}
	\label{eq:energykin}
	\mathbf{J}_{kin}^{E}(\mq=0)=\sum_{f}\sum_{\lambda}\int \frac{d^2k}{(2\pi)^2} \frac{ \mk k^2}{2(m)^2}  c^\dagger_{\lambda f}(\mk) c_{\lambda f}(\mk).
	\end{equation}
	In comparison with the charge current operator, there is no \textit{Zitterbewegung} contribution to the kinetic part of the energy current. We also see that quasi-particle and quasi-hole bands contribute to the energy current equally. At each momentum $\mathbf{k}$, quasi-particle and quasi-hole bands have the same energy and velocity and hence make the same contribution to the energy current .  
	\subsubsection{Interaction contribution}
	In the linear response calculation, the contribution of the Coulomb interaction to the energy density is given by 
	\begin{align}
	\delta \mathbf{H}^C(\mx)&\equiv\frac{\delta \hat{V}_C}{\delta \rho(\mathbf{x})}=N_0 \int d^2 Y V_C(|\mathbf{x}-\mathbf{Y}|)\delta \rho (\mx)\\&=N_0 V_C(\mq=0)\delta\rho(\mx)=N_0 \frac{2\pi \alpha}{q_{TF}} \delta \rho(\mx),
	\end{align}
	where $N_0$ is the total background charge number.

	The contribution to the energy current from the Coulomb interaction is then given by 
	\begin{equation}
	\label{eq:EnergyCoulombJ}
	\mathbf{J}^E_C(\mq=0)=N_0\frac{2\pi \alpha}{e q_{TF}}\mathbf{J}(\mq=0),
	\end{equation}
	where $\mathbf{J}(\mq=0)$ is nothing but the charge current. However in the kinetic formalism, we consider $\delta \mathbf{H}^C(\mx)$ as the shift of the chemical potential due to the background charge (the Hartree diagram). 
	\subsection{Stress tensor operator}
	The effective Lagrangian in curved space is defined as 
	\begin{equation}
	\label{eq:effactcur}
	\cS= \int dt \left(d^2 \x \sqrt{g(\x)} \sum_{\xi}\mathcal{L}_{\xi}(\x)-\hat{V}_C  \right),
	\end{equation} 
	where the free Lagrangian density $\mathcal{L}_{\xi}(\x)$ is defined in equation \eqref{eq:freeL} and $\hat{V}_C $ is the effective Coulomb interaction.
	The stress tensor is defined as the response of the system with respect to a perturbation of the local metric,
	\begin{equation}
	\label{eq:Tijdef}
	T^{ij}(\x)=-\frac{2}{\sqrt{g}(\x)}\frac{\delta \cS(g_{ij}(\x),\tilde{\hat{\Psi}}^\dagger,\tilde{\hat{\Psi}})}{\delta g_{ij}(\x)}\arrowvert_{\delta g_{ij}=0}
	\end{equation}
	where $g_{ij}(\x)=\delta_{ij}+\delta g_{ij}(\x)$ and the rescaled field is 
	\begin{equation}
	\label{eq:curvedpsi}
	\tilde{\hat{\Psi}}=g^{1/4}\hat{\Psi}.
	\end{equation}
	\subsubsection{Kinetic contribution}
	We calculate the stress tensor operator for the kinetic Hamiltonian \eqref{eq:Ham} following Ref \cite{Fujikawa1981}. We leave the detailed calculation to Appendix \ref{sec:operaotsApp}, where we derive the results directly using definition \eqref{eq:Tijdef} and the explicit form of the kinetic Hamiltonian in curved space.  Here we quote the result of the kinetic contribution to the stress tensor 
	\begin{equation}
	\label{eq:stresskin}
	T^{ij}(\mq=0)=T^{(I)ij}(\mq=0)+T^{(II)ij}(\mq=0),
	\end{equation} 
	where the normal contribution to the kinetic part of stress tensor is 
	\begin{align}
	\label{eq:Tijnormal}
	T^{(I)ij}(\mq=0)=\sum_{\xi}T^{(I)ij}_\xi(\mq=0)\nonumber\\=\sum_f \sum_\lambda\int \frac{d^2 k}{(2\pi)^2}\frac{\lambda k^i k^j}{m} c^\dagger_{\lambda f}(\mk) c_{\lambda f}(\mk).
	\end{align}
	Equation \eqref{eq:Tijnormal} has a similar form as the stress tensor operator for a quadratic semimetal \cite{Dumitrescu:2015} like HgTe. 
	
	The \textit{Zitterbewegung} contribution to the kinetic part of the stress tensor is given by
	
	\begin{multline}
	T^{(\text{II})11}(\mq=0)=-T^{(\text{II})22}(\mq=0)\\=i\sum_f \int \frac{d^2 k}{(2\pi)^2} \frac{k^1 k^2}{m}\left[c^\dagger_{+ f}(\mk) c_{- f}(\mk)-c^\dagger_{- f}(\mk) c_{+ f}(\mk)\right],
	\end{multline}

	\begin{multline}
	T^{(\text{II})12}(\mq=0)=T^{(\text{II})21}(\mq=0)=-i\sum_f \int \frac{d^2 k}{(2\pi)^2} \\\times\frac{ (k^1)^2 -(k^2)^2}{2m}\left[c^\dagger_{+ f}(\mk) c_{- f}(\mk)-c^\dagger_{- f}(\mk) c_{+ f}(\mk)\right].
	\end{multline}

	\subsubsection{Interaction contribution}
	Besides the kinetic contribution, the stress tensor also has a contribution from the interactions, which we will calculate now. We turn on the homogeneous metric perturbation $\delta g_{ij}$. The Coulomb interaction in curved space near charge neutrality is given by 
	
	\begin{equation}
	\label{eq:Coulombcurved}
	\hat{V}_C=\frac{1}{2\sqrt{g}}\int \frac{d^2 p}{(2\pi)^2} \frac{2\pi \alpha}{|\mathbf{p}|+q_{TF}}\tilde{\rho}^{\text{eff}}(-\mathbf{p})\tilde{\rho}^{\text{eff}}(\mathbf{p}),
	\end{equation}
	where $\tilde{\rho}^{\text{eff}}(\mq)$ is given by substituting 
	\begin{equation}
	\label{eq:curvedc}
	c^\dagger_{\lambda f}\rightarrow g^{1/4} c^\dagger_{\lambda f}, \qquad c_{\lambda f}\rightarrow g^{1/4} c_{\lambda f},
	\end{equation}
	in \eqref{eq:rho_eff}, and $q=|\mq|=\sqrt{g^{ij}q_iq_j}$.
	The factor $g^{-1/2}$ appears in equation \eqref{eq:Coulombcurved} as was shown in the previous section. The transformation \eqref{eq:curvedc} is equivalent to the transformation \eqref{eq:curvedpsi}. We now can use the definition of the stress tensor \eqref{eq:Tijdef} to derive the contribution of the Coulomb interaction to the stress tensor in flat space time by taking the derivative of $\hat{V}_C$ with respect to the homogeneous metric \footnote{There is a metric dependence in the definition of $|\mathbf{p}|$.}
	\begin{align}
	\label{eq:TC}
	T_C^{ij}(\mq=0)=&\pi \alpha \int \frac{d^2 p}{(2\pi)^2}\Big[-\frac{p^i p^j}{p}\frac{1}{(p+q_{TF})^2}+\delta^{ij}\frac{1}{p+q_{TF}}\Big]\nonumber\\&\times\rho^{\text{eff}}(\mathbf{p})\rho^{\text{eff}}(-\mathbf{p}).
	\end{align}
	We leave the detailed derivation of \eqref{eq:TC} to Appendix \ref{sec:operaotsApp}. The contribution to $T^{ij}_C(\mq=0)$ up to linear order in the perturbation is given by\footnote{In this paper, we are only interested in the linear transport calculation.} 
	\begin{align}
	T_C^{ij}(\mq=0)&=2\pi\alpha \frac{\delta^{ij}}{q_{TF}(0)}N_0 \rho^{\text{eff}}(0)\\&=2\pi \alpha N_0 \sum_f \sum_{\lambda} \int \frac{d^2 k}{(2\pi)^2} \frac{\delta^{ij}}{q_{TF}(0)} c^\dagger_{\lambda f}(\mk) c_{\lambda f}(\mk), \nonumber
	\end{align}
	where $N_0$ is the background charge. We can view this contribution simply as a shift in the chemical potential. In the calculation for the shear stress tensor in section \ref{sec:Kinetic_Equation}, we will calculate $T^{12}$ under a constant shear. The contribution from the interactions $T^{ij}_C \sim \delta^{ij}$ will hence not enter our calculation.

	\section{Kinetic equation and quantum transport}
	\label{sec:Kinetic_Equation}
	After having derived the conserved currents, we are now ready to begin the derivation of the actual QBE. In this vein, we will set up the semi-classical problem of electron and hole transport in bilayer graphene at a finite temperature $T$. We define the retarded Green's function as follows \footnote{We ignore the flavor index $f$.}  
	\begin{multline}
	\label{eq:Green}
	g^{<}_{\lambda \lambda'} (\mathbf{k},\omega,\mathbf{x},t)=i\int d^2 \mathbf{r}d\tau e^{i(\omega\tau-\mathbf{k}\cdot \mathbf{r})}\\\times\<\Psi^{\dagger}_{\lambda'}(\mathbf{x}-\frac{\mathbf{r}}{2},t-\frac{\tau}{2})\Psi_\lambda(\mathbf{x}+\frac{\mathbf{r}}{2},t+\frac{\tau}{2})\>,
	\end{multline}
	where $\lambda$ and $\lambda'$ are the band indices. The expectation value $\langle \,\, \rangle$ is evaluated at finite temperature as explained in more detail in Appendix \ref{sec:BoltzA}. In order to study the DC transport, we can ignore the off-diagonal part of the retarded Green's function since this part depends explicitly on time as explained in Appendix \ref{sec:BoltzA}. Equation \eqref{eq:Green} now takes the explicit form
	\begin{equation}
	\label{eq:gdiag}
	g^{<}_{\lambda \lambda'} (\mathbf{k},\omega,\mathbf{x},t)=2\pi i\delta(\omega-\epsilon_{\lambda}(\mathbf{k}))f_{\lambda}(\mathbf{k},\mathbf{x},t)\delta_{\lambda {\lambda'}},
	\end{equation}
	where $f_{+(-)}(\mathbf{p},\mathbf{x},t)$ is the distribution function of electrons in the conduction (valence) band.  We can write down formally the QBE for the distribution function as 
	\begin{multline}
	\label{eq:Boltz1}
	\left(\frac{\p}{\p t}+\mathbf{v}_\lambda(\mathbf{k}) \cdot\frac{\p}{\p \x} +e\mathbf{E}(\mathbf{x},t)  \cdot \frac{\p}{\p \mk}\right)f_{\lambda}(\mk,\x,t)\\=-I_{\lambda}[\{f_{\lambda_i}\}](\mk,\mathbf{x},t),
	\end{multline}
	where the group velocity of band $\lambda$ is defined as 
	\begin{equation}
	\mathbf{v}_{\lambda}(\mathbf{k})=\p_{\mathbf{k}}\epsilon_{\lambda}(\mathbf{k}).
	\label{eq:v_group}
	\end{equation}
	$\mathbf{E}(\mathbf{x},t)$ is slowly varying applied electric field. The right-hand side of the equation is the collision integral, which can be derived explicitly from first principles. In section \ref{sec:Coll}, we will discuss in detail the collision integral, which takes into account the scattering of quasi-particles off each other, on impurities as well as at the boundary. The microscopic derivation of equation \eqref{eq:Boltz1} is left for Appendix \ref{sec:BoltzA}. In the subsequent subsections, we will employ the equation \eqref{eq:Boltz1} to set up the calculation of the transport coefficients. In DC transport, we can ignore the contribution from the \textit{Zitterbewegung} contribution which comes from the off-diagonal part of the Green's function \eqref{eq:Green}. From the results \eqref{eq:chargecurrent1}, \eqref{eq:heat} with \eqref{eq:energykin},  and \eqref{eq:Tijnormal} in the previous section, we can obtain expressions for the expectation value of the normal contribution to the conserved currents in terms of the local distribution function as follows
	\begin{equation}
	\label{eq:currentexpect}
	  \mathbf{J}=  N_f\frac{e}{m}\sum_{\lambda}\int \frac{d^2k}{(2\pi)^2} \lambda \mk f_{\lambda}(\mk),
	\end{equation}

		\begin{equation}
	\label{eq:Qcurrentexpect}
	  \mathbf{J}^{Q}=  N_f\sum_{\lambda}\int \frac{d^2k}{(2\pi)^2} \frac{\lambda \mk}{m}(\epsilon_{\lambda}(\mk)-\mu) f_{\lambda}(\mk),
	\end{equation}
	and the kinetic contribution to the stress tensor has the expectation value 
		\begin{align}
	\label{eq:Tijexpect}
	T^{ij}=N_f \sum_\lambda\int \frac{d^2 k}{(2\pi)^2}\frac{\lambda k^i k^j}{m} f_{\lambda}(\mk).
	\end{align}
	We note that the results in Eqns.~\eqref{eq:currentexpect}-\eqref{eq:Tijexpect} look similar to the Fermi liquid results for two types of particles, although we are in a very different regime without a well-formed Fermi surface. If we replace the distribution function $f_{\lambda}(\mk)$ in equations \eqref{eq:currentexpect},  \eqref{eq:Qcurrentexpect} and \eqref{eq:Tijexpect} by the unperturbed Fermi distribution $f^0_\lambda(\mk)$, we get zero. In this section, we will use the above equations to obtain the expectation value of the conserved currents in terms of the distribution function perturbations:
		\begin{equation}
	f_\lambda(\mk,\mathbf{x})=f^0_\lambda(\mk)+f^0_\lambda(\mk)[1-f^0_\lambda(\mk))]h_\lambda(\mk,\mathbf{x}).
	\label{eq:f_pert}
	\end{equation}
	\subsection{Constant applied magnetic field}
	\label{sec:Bfield}
	So far, the experiments performed on the electrical conductivity of suspended BLG have been performed in zero magnetic field. However, we believe it is eminently possible to extend the experiments in this direction and to this end, we will set up the calculation process to obtain the transport coefficients with an applied magnetic field $\mathbf{B}=B\hat{\mathbf{z}}$. In order to use the kinetic equation with a magnetic field, we need to consider a weak magnetic field. In a Fermi liquid at zero temperature, the requirement is $k_F \ell_B \gg 1$ where the magnetic length is given by $\ell_B=\sqrt{\frac{\hbar c}{e B}}$. For neutral BLG, at finite temperature, one may guess that the valid limit of the kinetic equation is $k_T \ell_B \gg 1$,	where the thermal momentum is defined as $k_T=\sqrt{2 k_B T m/\hbar^2}$.
	For temperature $T=10K$, the appropriate magnetic field is $B<100$ Gauss. Such a small magnetic field also guarantees that the Zeeman energy term is small enough, that we can neglect the energy different between the two spin species.
	
	With the appearance of a magnetic field, we need to add one more term in the left-hand side of the kinetic equation to take into account the Lorentz force \cite{Maciejko2007}\footnote{We ignore the Zeeman effect, since it is small in comparison with the experimental temperature \cite{Nam2017}.}
	\begin{equation}
	\label{eq:B}
	e\left[\mathbf{v}_{\lambda}({\mathbf{k}})\times \mathbf{B} \right]\cdot \nabla_{\mathbf{k}} f_{\lambda}(\mathbf{k}, \mathbf{x},t)
	\end{equation}
	where the group velocity is given by \eqref{eq:v_group}.
	In this section, we only consider the charge conductivity and thermal conductivity in the appearance of a magnetic field. We can rewrite \eqref{eq:B} as 
	\begin{equation}
	\label{eq:ExtraB}
	-\frac{eB\lambda}{m}f^0_\lambda(\mathbf{k})[1-f^0_\lambda(\mathbf{k})]\epsilon^{ij}k_j \partial_{k_i}h_\lambda(\mathbf{k},\mathbf{x}).
	\end{equation}

	\subsection{Thermoelectric coefficients}
	We define the electrical conductivity $\sigma$, the thermal conductivity $K$ and the thermopower $\Theta$ by
	\begin{equation}
	\begin{pmatrix}
	\vec{J}\\
	\vec{J}^Q
	\end{pmatrix}
	=\begin{pmatrix}
	\sigma &\Theta\\
	T\Theta&K
	\end{pmatrix}
	\begin{pmatrix}
	\vec{E}\\
	-\vec{\nabla}T
	\end{pmatrix}
	\label{eq:def_thermo}
	\end{equation}
	where each of the thermoelectric coefficients is a $2\times2$ matrix. The fact that $\Theta$ appears twice in \eqref{eq:def_thermo} is due to the Onsager reciprocity relation \cite{Onsager1931}. Using these definitions, the Seebeck coefficient is $S=\sigma^{-1}\Theta$ and the Peltier coefficient is $\Pi=TS$. In experiments, the heat current is often measured such that $\vec{J}=0$, in which case the proportionality constant between $\vec{J}^Q$ and $-\vec{\nabla}T$ is $\kappa=K-T\Theta\sigma^{-1}\Theta$ \cite{Mueller2008,lucas2018hydrodynamics}.

	\subsection{Charge conductivity}
	In order to derive the coefficients of DC conductivity, we apply a constant electric field $\mathbf{E}$. The unperturbed distribution function is given by 
	\begin{equation}
	f^0_\lambda(\mathbf{p})=\frac{1}{1+e^{\beta(\epsilon_\lambda(\mathbf{p})-\mu)}}.
	\label{eq:f_0}
	\end{equation}
	We need to solve the equation \eqref{eq:Boltz1} in the following simplified form
	\begin{align}
	\label{eq:BolE}
	-\lambda& \beta\frac{e\mathbf{E}\cdot\mathbf{p}}{m} f^0_\lambda(\mathbf{p})[1-f^0_\lambda(\mathbf{p})]\nonumber\\+&\frac{eB\lambda}{m}f^0_\lambda(\mathbf{p})[1-f^0_\lambda(\mathbf{p})]\epsilon^{ij}p_j \nabla_{p_i}h_\lambda(\mathbf{p})\\&=-I^{(1)}_\lambda[\{h_{\lambda_i}(\mk_i)\}](\mathbf{p})\nonumber
	\end{align}
	for $\lambda=+$ and $\lambda=-$ to obtain $h_\lambda(\mathbf{p})$. In equation \eqref{eq:BolE}, the right-hand side denotes the linear order in the perturbation of the collision integral derived in section \ref{sec:Coll}. The left-hand side is derived in the Green's function formalism as \eqref{eq:KineticBandL1linear1}. The suggested ansatz in this calculation is 
	\begin{equation}
	\label{eq:as12}
	h_\lambda(\mathbf{p})=\beta\frac{e\mathbf{E}}{m}\cdot\bigg(\mathbf{p}\chi_\lambda^\parallel(p)+\mathbf{p}\times\hat{\vec{z}}\chi_\lambda^\perp(p)\bigg)
	\end{equation}
	and we solve for $\chi_\lambda(p)$ numerically. The second term in the ansatz becomes relevant when we have a magnetic field. The charge current is given by \eqref{eq:currentexpect} and the DC conductivity can be directly read off. Due to the symmetry of the collision integral that will be discussed in section \ref{sec:sym}, we can show that $\sigma_{xx}=\sigma_{yy}$ because of rotational invariance and $\sigma_{xy}=\sigma_{yx}=0$ in the absence of magnetic field due to parity. The external magnetic field $\mathbf{B}$ breaks parity which gives us $\sigma_{xy}=-\sigma_{yx} \neq 0$. 

	\subsection{Thermal conductivity}
	
	We consider a spatially dependent background temperature $T(\mathbf{x})=T+\delta T(\mathbf{x})$. The local  equilibrium distribution function takes the form 
	\begin{equation}
	f^0_\lambda(\mathbf{p},T(\mathbf{x}),\mu)=\frac{1}{1+e^{\frac{1}{k_BT(\mathbf{x})}\left(\epsilon_\lambda(\mathbf{p}) -\mu\right)}}.
	\end{equation} 
	We now consider a constant gradient in temperature by introducing the space-time independent driving \textit{force}
	$\mathbf{F}^T=-\frac{\nabla_{\mathbf{x}}\delta T}{T}$.  We then need to solve equation  \eqref{eq:Boltz1} in the following simplified form
	\begin{align}
	-\lambda& \beta\frac{\mathbf{F}^T\cdot\mathbf{p}}{m}(\epsilon_\lambda(\mathbf{p})-\mu) f^0_\lambda(\mathbf{p})[1-f^0_\lambda(\mathbf{p})]\nonumber\\+&\frac{eB\lambda}{m}f^0_\lambda(\mathbf{p})[1-f^0_\lambda(\mathbf{p})]\epsilon^{ij}p_j \nabla_{p_i}h_\lambda(\mathbf{p})\\&=-I^{(1)}_\lambda[\{h_{\lambda_i}(\mk_i)\}](\mathbf{p})\nonumber
	\end{align}
	for $\lambda=+$ and $\lambda=-$ to obtain $h_\lambda(\mathbf{p})$. The left-hand side is obtained from \eqref{eq:KineticBandL4linear3}. The suggested ansatz is 
	\begin{equation}
	\label{eq:as32}
	h_\lambda(\mathbf{p})=\beta\frac{\mathbf{F}^T}{m}(\epsilon_\lambda(\mathbf{p})-\mu)\cdot\bigg(\mathbf{p}\phi_\lambda^\parallel(p)+\mathbf{p}\times\hat{\vec{z}}\phi_\lambda^\perp(p)\bigg).
	\end{equation}
From the heat current along with equations \eqref{eq:Qcurrentexpect} we can read off the thermal conductivity. For the thermopower, we consider the same ansatz as for the thermal conductivity, but calculate the charge current which is given by \eqref{eq:currentexpect}.

	\subsection{Viscosity} 
	To calculate the DC shear viscosity, we consider a background velocity for the particles and holes. Therefore, the local equilibrium distribution function takes the form
	\begin{equation}
	f^0_\lambda(\mathbf{p},\mathbf{u}_\lambda(\mathbf{x}),\mu)=\frac{1}{1+e^{\beta(\epsilon_\lambda(\mathbf{p})-\mathbf{u}_\lambda(\mathbf x)\cdot\mathbf{p}-\mu)}},
	\end{equation}
	where $\mathbf{u}_{+(-)}(\x)$ is the perturbed background velocity of electrons (holes).
	We apply a constant shear with the explicit form
	\begin{equation}
	u^\lambda_{ 12}=F_\lambda, \qquad u^{\lambda}_{11}= u^{\lambda}_{22}=0
	\label{eq:strain}
	\end{equation}  
	where $F_\lambda$ is a space-time independent perturbation and the definition of strain is 
	\begin{equation}
		\label{eq:strain1}
	u^\lambda_{ ij}=\frac{1}{2}\left( \partial_i u_{\lambda}^j+\partial_j u_{\lambda}^i\right).
	\end{equation}
	We need to solve equation  \eqref{eq:Boltz1} in the following simplified form
	\begin{equation}
	\lambda \beta\frac{2p^1p^2F_\lambda}{m} f^0_\lambda(\mathbf{p})[1-f^0_\lambda(\mathbf{p})]=-I^{(1)}_\lambda[\{h_{\lambda_i}(\mk_i)\}](\mathbf{p})
	\end{equation}
	for $\lambda=+$ and $\lambda=-$ to obtain $h_\lambda(\mathbf{p})$. The left-hand side comes from \eqref{eq:KineticBandL3Linear2}. The suggested ansatz is 
	\begin{equation}
	\label{eq:as2}
	h_\lambda(\mathbf{p})=\beta\frac{2p^1 p^2}{m}\chi^\lambda_{\eta}(\mathbf{p};F_+,F_-).
	\end{equation}
	The stress tensor $T^{12}_\lambda$ is given by \eqref{eq:Tijexpect} and the shear viscosity coefficients are given by the definition 
	\begin{equation}
	T^{12}_\lambda=-\eta_{\lambda \lambda'}F_{\lambda'}.
	\end{equation}
	In the experiment \cite{Kumar2017}, the authors found that quasi-particle collisions can  importantly impact the transport in monolayer graphene. The results showed that the electrons behave as a highly viscous fluid due to the electron-electron interactions in the clean limit. Even though there has not been an analogous experiment for BLG yet, we expect that highly viscous behaviour of BLG will be found in the near future. The viscosity coefficients will play an important role for simulation of electronic transport in BLG and for comparison against experimental results.  
	\\
	\\
	
	\section{Collision integral}
	\label{sec:Coll}
	We now focus on the right-hand side of the QBE---the collision integral. We discuss the contribution from quasi-particle interactions, scattering on disorder and scattering off the boundary separately in subsections \ref{subsec:CI}, \ref{subsec:DI} and \ref{subsec:BI} respectively. In the quasi-particle scattering channel, we ignore Umklapp processes at low energies near charge neutrality. Since in our regime $k_F a\ll 1$, Umklapp scattering is negligible due to the lack of available phase space. Inter-valley scattering is also ignored due to the long range nature of the Coulomb interaction. Up to linear order in the perturbation, the generalized collision integral on the right-hand side of the kinetic equation \eqref{eq:Boltz1} includes contributions from quasi-particle interactions, scattering on disorders and finite size effect and scattering on phonons respectively
	\begin{equation}
	\label{eq:CoItot}
	I^{(1)}_{\lambda}=I^{(1)}_{\lambda,\textrm{int}}+I^{(1)}_{\lambda,\textrm{dis}}+I^{(1)}_{\lambda,\textrm{size}}+I^{(1)}_{\lambda,\textrm{phonon}}.
	\end{equation}
	In the following subsection, we will discuss in detail each contribution of \eqref{eq:CoItot}.
	\subsection{Quasi-particles' Coulomb interaction}
	
	\label{subsec:CI}
	The first contribution to the collision integral that we consider is that coming from the Coulomb interaction of the quasi-particles. We are interested in the experimental regime of sufficiently clean BLG \cite{Nam2017} in which the transport properties are dominated by quasi-particle interactions. In this section, we formulate the quasi-particle interactions of the form \eqref{eq:Coulomb} via the screened Coulomb potential $V_C(\mathbf{q})$ that was derived previously in section \ref{sec:Coul}. To derive the contribution $I^{(1)}_{\lambda,\textrm{int}}$, we generalize the Kadanoff-Baym  equations \cite{Kadanoff} to BLG. We again only consider the diagonal component of the Green's functions \eqref{eq:gdiag} and calculate the collision integral contribution due to the interaction \eqref{eq:Coulomb}. The technical details of the derivation will be left for Appendix \ref{sec:BoltzA}, in this subsection we only quote the main result. The collision integral due to interactions for each band index $\lambda$ is then given by
	\begin{widetext}
		\begin{multline}
		\label{eq:CoI}
		I_{\lambda,\textrm{int}}[\{f_{\lambda_i}(\mk_i)\}](\mathbf{p})=-(2\pi)\sum_{\lambda_1\lambda_2\lambda_3}\int \frac{d^2 \mk_1}{(2\pi)^2}\frac{d^2 \mq}{(2\pi)^2}\delta(\epsilon_\lambda(\mathbf{p})+\epsilon_{\lambda_1}(\mk_1)-\epsilon_{\lambda_2}(\mathbf{p}+\mq)-\epsilon_{\lambda_3}(\mk_1-\mq))\\
		\left[N_f|T_{\lambda\lambda_1\lambda_3\lambda_2}(\mathbf{p},\mk_1,\mq)|^2-T_{\lambda\lambda_1\lambda_3\lambda_2}(\mathbf{p},\mk_1,\mq)T^*_{\lambda\lambda_1\lambda_2\lambda_3}(\mathbf{p},\mk_1,\mk_1-\mathbf{p}-\mq)\right]\\
		\Big[[1-f_\lambda(\mathbf{p})][1-f_{\lambda_1}(\mk_1)]f_{\lambda_2}(\mathbf{p}+\mq)f_{\lambda_3}(\mk_1-\mq)-f_\lambda(\mathbf{p})f_{\lambda_1}(\mk_1)[1-f_{\lambda_2}(\mathbf{p}+\mq)][1-f_{\lambda_3}(\mk_1-\mq)]\Big],
		\end{multline}
	\end{widetext} 
	where we follow \cite{Fritz2008} and define the form factor 
	\begin{equation}
	M_{\lambda\lambda'}(\mk,\mk')=\frac{1}{2}\left(1+\lambda\lambda'e^{i(2\theta_{\mk'}-2\theta_{\mk})}\right),
	\end{equation}
	as well as the channel dependent scattering matrix
	\begin{equation}
	T_{\lambda_1\lambda_2\lambda_3\lambda_4}(\mk,\mk',\mq)=V(-\mq)M_{\lambda_1\lambda_4}(\mk+\mq,\mk)M_{\lambda_2\lambda_3}(\mk'-\mq,\mk').
	\end{equation}
	The collision integral vanishes when we substitute the Fermi distribution \eqref{eq:f_0}.
    Linear order in perturbation of the collision integral is given by the perturbation \eqref{eq:f_pert}.
	The linearized collision integral is then given by 
	\begin{widetext}
		\begin{multline}
		I_{\lambda,\textrm{int}}^{(1)}[\{h_{\lambda_i} (\vec{k}_i)\}](\vec{p})=-(2\pi)\sum_{\lambda_1\lambda_2\lambda_3}\int \frac{\diff^2\vec{k}}{(2\pi)^2} \frac{\diff^2\vec{q}}{(2\pi)^2}\delta(\epsilon_{\lambda}(\vec{p})+\epsilon_{\lambda_1}(\vec{k})-\epsilon_{\lambda_2}(\vec{p+q})-\epsilon_{\lambda_3}(\vec{k-q}))\\
		\times\bigg[N_f|T_{\lambda\lambda_1\lambda_3\lambda_2}(\vec{p},\vec{k},\vec{q})|^2-T_{\lambda\lambda_1\lambda_3\lambda_2}(\vec{p},\vec{k},\vec{q})T_{\lambda\lambda_1\lambda_2\lambda_3}^*(\vec{p},\vec{k},\vec{k-p-q})\bigg]\\
		\times \bigg[[1-f_\lambda^0(\vec{p})][1-f_{\lambda_1}^0(\vec{k})]f_{\lambda_2}^0(\vec{p+q})f_{\lambda_3}^0(\vec{k-q})\bigg]\bigg[-h_{\lambda}(\vec{p})-h_{\lambda_1}(\vec{k})+h_{\lambda_2}(\vec{p+q})+h_{\lambda_3}(\vec{k-q})\bigg],
		\label{eq:lin_coll_int}
		\end{multline}
	\end{widetext}
	where we defined $h_\lambda(\mk)$ in equation \eqref{eq:f_pert}. The collision integral \eqref{eq:lin_coll_int} shares similarities with one of monolayer graphene in Ref \cite{Fritz2008}. However, due to the difference in the quasi-particle dispersion relation, which is quadratic for BLG and linear for monolayer graphene, their allowed scattering channels differ qualitatively. In the case of BLG, we have to consider the scattering channel where one quasi-particle decays to two quasi-particles and one hole. On the other hand, this scattering channel is kinematically forbidden in monolayer graphene because of momentum and energy conservation. This contribution was missed in a previous publication on the kinetic theory of BLG \cite{Lux2013}. However, due to kinematic restrictions, the phase space for this scattering process is small and therefore this channel does not contribute significantly to the collision integral. We have checked this statement numerically.

	\subsection{Contribution from disorder}
	\label{subsec:DI}
	Due to the Galilean invariance of our system in the absence of disorder, the collision integral is unchanged under a Galilean boost. However, under a uniform boost of all particles by
$\vec{u}$, the current density transforms as $	\vec{J}\to\vec{J}+en\vec{u}$. So as long as the charge density $n\neq0$ (ie $\mu\neq0$) we change the current density by boosting frames and therefore the conductivity is ill-defined in the absence of a momentum-relaxing mechanism. Including one or several momentum-relaxing scattering channels is therefore crucial for calculating the transport coefficients away from $\beta\mu=0$. 

One such momentum relaxing process is the scattering of electrons off impurities in the sample. For this calculation, we put our system in a box of side length $L$ with periodic boundary conditions. We follow \cite{Fritz2008} and consider a disorder Hamiltonian
	\begin{equation}
	H_{\textrm{dis}}=\sum_f\int d^2\vec{x}\ V_{\textrm{dis}}(\vec{x})\Psi_f^\dagger(\vec{x})\Psi_f(\vec{x}),
	\label{eq:H_dis}
	\end{equation}
	where $V_{\textrm{dis}}$ is the interaction potential between an electron and the impurities, which we take to be charges $Ze$ located at random positions $\vec{x}_i$ and having number density $n_{\textrm{imp}}=N_{\textrm{imp}}/L^2$. We use the screened Coulomb interactions to obtain
	\begin{equation}
	V_{\textrm{dis}}(\vec{x})=\sum_{i=1}^{N_{\textrm{imp}}}\frac{Ze^2}{\epsilon_r|\vec{x}-\vec{x}_i|}e^{-q_{TF}|\vec{x}-\vec{x}_i|}.  
	\end{equation}
	From the interaction \eqref{eq:H_dis}, we can calculate the scattering rate of quasi-particles off the disorder. We then obtain the contribution to the collision integral from disorder up to linear order in the perturbation 
	\begin{equation}
	\label{eq:CoD}
	I^{(1)}_{\lambda,\textrm{dis}}[h_{\lambda_i} (\vec{k}_i)](\vec{p})=\tau_\textrm{imp}^{-1} f_\lambda^0(p)[1-f_\lambda^0(p)]h_\lambda(\vec{p}),
	\end{equation}
	where we define a short hand notation for the impurity scattering rate
	\begin{equation}
	\label{eq:tau_imp}
	\tau_\textrm{imp}^{-1}=\frac{1}{2}mn_{\textrm{imp}}\bigg(\frac{2\pi Ze^2}{\epsilon_rq_{TF}}\bigg)^2.
	\end{equation}
	The corresponding dimensionless parameter is $\alpha_\textrm{imp}\equiv \beta\tau_\textrm{imp}^{-1}=1/2(8\pi^2Z/N_f\epsilon_r)^2\beta n_{\textrm{imp}}/m$.
	The detailed derivation of \eqref{eq:CoD} is left for Appendix \ref{sec:CoD}.
	\subsection{Effect of finite system size}
	\label{subsec:BI}
	In very clean samples of bilayer graphene, it is expected that the scattering length due to impurity scattering is longer than the system size $L$, which is currently limited in suspended graphene samples \cite{Ki2013}. In this case, in order to have a well-defined conductivity, we need to include the effect of the finite size of the system. There will be scattering of the electrons off the boundary, which effectively acts as an additional scattering time. Assume the scattering time due to collisions with the boundary is $	\tau(p)=\frac{L}{v}=\frac{Lm}{p}$ where $L$ is the size of the sample up to a factor depending on the geometry of the BLG sample. Here, we are making the simplifying assumption, that that the scattering does not depend on the direction of the momentum. We neglect the angular dependence of the boundary scattering which is likely to contribute a geometric factor to the scattering time. The collision integral is then
	\begin{equation}
	\label{eq:CoS}
	I^{(1)}_{\lambda,\textrm{size}}[h_{\lambda_i} (\vec{k}_i)](\vec{p})=\frac{p}{mL}f_\lambda^0(p)[1-f_\lambda^0(p)]h_\lambda(\vec{p})
	\end{equation}
	which is just the Bhatnagar-Gross-Krook (BGK) collision operator \cite{BGK} with $\tau$ given by $\tau(p)$. The corresponding dimensionless parameter is $\alpha_L=\frac{\sqrt[]{\beta}}{\sqrt[]{m}L}$.

    \subsection{Phonon scattering}
    We should also consider the effect of the electrons scattering off phonons. The maximum energy of an acoustic phonon is $\varepsilon_{\textrm{max}}=2c\sqrt{2m\textrm{max}(k_BT,\mu)}$, where $c$ is the speed of sound in graphene. In the experimental setting, we are at high temperatures compared to the Bloch-Gr\"uneisen temperature
    \begin{equation}
        T_{BG}=\frac{2c}{v_F}\frac{\sqrt{\gamma_1|\mu|}}{k_B},
    \end{equation}
     and additionally we have $T\gg 2\gamma_1/k_B(c/v)^2$. Thus we are in the high temperature regime $k_BT\gg\varepsilon_{\textrm{max}}$, where can treat the phonons as introducing another scattering time \cite{Viljas2010}\footnote{We ignore the scattering channel from conductance  electrons to valence electrons due to the emission and absorption of phonons because of the suppression in the scattering matrix.\cite{Viljas2010}.}
    \begin{equation}
        \tau_\textrm{phonon}^{-1}=\frac{D^2mk_BT}{2\rho\hbar^3
        c^2},
        \label{eq:tau_phonon}
    \end{equation}
where $D$ is the deformation potential and $\rho$ is the mass density. Then the collision integral is
\begin{equation}
	I^{(1)}_{\lambda,\textrm{phonon}}[h_{\lambda_i} (\vec{k}_i)](\vec{p})=\tau_\textrm{phonon}^{-1}f_\lambda^0(p)[1-f_\lambda^0(p)]h_\lambda(\vec{p}).
	\label{eq:I_phonon}
	\end{equation}
	The corresponding dimensionless parameter is $\alpha_{\textrm{ph}}=\beta \tau_\textrm{phonon}^{-1}$. It is crucial to note that whereas $\alpha_{\textrm{imp}}=\alpha_{\textrm{imp}}(T)$ and $\alpha_{\textrm{L}}=\alpha_{\textrm{L}}(T)$, $\alpha_{\textrm{ph}}$ does not depend on temperature.

		\section{Symmetries}
	\label{sec:sym}
	\subsection{Spatial symmetries}
	The electrical conductivity is rotationally symmetric. The only rotationally symmetric tensors in 2d are $\delta_{ij}$ and $\epsilon_{ij}$ so any rotationally invariant tensor $\sigma_{ij}$ can be written as
	\begin{equation}
	\sigma_{ij}=\sigma_{xx}\delta_{ij}+\sigma_{xy}\epsilon_{ij}.
	\end{equation}
	In the absence of a magnetic field, we have an additional symmetry, namely 2D parity $y\to-y$, which implies 
	\begin{equation}
	\sigma_{xy}=\sigma_{yx}=0.
	\end{equation}
	With a magnetic field, 2D parity implies
	\begin{align}
	    \sigma_{xx}(B)&=\sigma_{xx}(-B),\\
	    \sigma_{xy}(B)&=-\sigma_{xy}(-B).
\end{align}
The thermal conductivity and thermopower satisfy the same relations.
	
	\subsection{Particle-hole symmetry}
    Under the particle-hole transformation, we have $\lambda\to-\lambda$, $\mu\to-\mu$ and $B\to-B$. First consider the electrical conductivity. Due to the particle-hole symmetry, we have
    \begin{align}
      \sigma_{xx}(\beta\mu,B)&=\sigma_{xx}(-\beta\mu,-B),\\ \sigma_{yx}(\beta\mu,B)&=\sigma_{yx}(-\beta\mu,-B).     
    \end{align}
 Now consider the viscosity which we only calculate for $B=0$. Particle-hole symmetry implies that we have
 \begin{equation}
     \eta_{\lambda,\lambda'}(\beta\mu)=\eta_{-\lambda,-\lambda'}(-\beta\mu).
 \end{equation}
These symmetries follow directly from the form of the collision integral \eqref{eq:lin_coll_int}.

\section{Two-fluid model}
\label{sec:2fm}
We introduce the two-fluid model, which reproduces the salient features of our numerical results. Motivated by comparison with experiment \cite{Wagner} we choose to only include the phonon scattering as a momentum relaxing mechanism. We multiply the kinetic equation by $\lambda \vec{p}/m$ and integrate over momentum space in order to derive the evolution of the mean fluid velocities as 
\begin{equation}
    m\partial_t\vec{u}^e=-\frac{m}{\tau_{eh}}(\vec{u}^e-\vec{u}^h)-\frac{m\vec{u}^e}{\tau_{se}}+e(\vec{E}+\vec{u}^e\times\vec{B})-\Lambda^e k_B\nabla T
    \label{eq:fluid1}
\end{equation}

\begin{equation}
    m\partial_t\vec{u}^h=\frac{m}{\tau_{he}}(\vec{u}^e-\vec{u}^h)-\frac{m\vec{u}^h}{\tau_{sh}}-e(\vec{E}+\vec{u}^h\times\vec{B})-\Lambda^h k_B\nabla T,
    \label{eq:fluid2}
\end{equation}
where we defined the electron and hole velocities as 
\begin{equation}
\mathbf{u}^e=\frac{\int \frac{d^2 \vec{p}}{(2\pi)^2}\frac{\mathbf{p}}{m} f_+(\mathbf{p})}{\int \frac{d^2 \vec{p}}{(2\pi)^2} f^0_+(\mathbf{p})},\quad \mathbf{u}^h=-\frac{\int \frac{d^2 \vec{p}}{(2\pi)^2}\frac{\mathbf{p}}{m} (1-f_-(\mathbf{p}))}{\int \frac{d^2 \vec{p}}{(2\pi)^2} (1-f^0_-(\mathbf{p}))}.
\end{equation}
The coefficients $\Lambda^{e,h}$ account for the fact that the average entropy per particle is $\Lambda k_B$ 

\begin{equation}
\label{eq:Lambdae}
    k_BT\Lambda^e=\frac{\int \frac{d^2 \vec{p}}{(2\pi)^2} p^2(\epsilon_+(p)-\mu)f^0_+(\mathbf{p})[1-f^0_+(\mathbf{p})]}{\int \frac{d^2 \vec{p}}{(2\pi)^2} f^0_+(\mathbf{p})}
\end{equation}

\begin{equation}
\label{eq:Lambdah}
    k_BT\Lambda^h=\frac{\int \frac{d^2 \vec{p}}{(2\pi)^2} p^2(-\epsilon_-(p)+\mu)f^0_-(\mathbf{p})[1-f^0_-(\mathbf{p})]}{\int \frac{d^2 \vec{p}}{(2\pi)^2}(1-f^0_-(\mathbf{p}))}
\end{equation}
The definitions \eqref{eq:Lambdae} and \eqref{eq:Lambdah} follow from the $\nabla T$ term in the QBE when \eqref{eq:fluid1} and \eqref{eq:fluid2} are derived. The Coulomb drag term can be derived explicitly from the collision integral
\begin{multline}
\int \frac{d^2 \vec{p}}{(2\pi)^2} \frac{\lambda\vec{p}}{m}I_{\lambda,\textrm{int}}^{(1)}\bigg[h_{\lambda_i} (\vec{k}_i)=\lambda_i\beta \vec{k}_i\cdot \bigg(\frac{\vec{u}^e-\vec{u}^h}{2}\bigg)\bigg](\vec{p})\\=
\left\{\begin{array}{ll}-\frac{mn^e}{\tau_{eh}}(\vec{u}^e-\vec{u}^h)   & \lambda=+ \\ 
\frac{mn^h}{\tau_{he}}(\vec{u}^e-\vec{u}^h) & \lambda=-
\end{array} \right. 
\label{eq:Coul_drag}
\end{multline} 
This allows us to calculate $\tau_{eh}$ and $\tau_{he}$. We perform the calculation at charge neutrality and then use \eqref{eq:away_from_CN} to extrapolate. $\tau_{se}$ is the momentum-relaxing scattering time for electrons and $\tau_{sh}$ is the corresponding time for holes ($s$ stands for "scattering"). They are given by
\begin{equation}
    \tau_{se}^{-1}=\tau_{\textrm{phonon}}^{-1}+\tau_{\textrm{imp}}^{-1}+\tau_{Le}^{-1},
\end{equation}
\begin{equation}
    \tau_{sh}^{-1}=\tau_{\textrm{phonon}}^{-1}+\tau_{\textrm{imp}}^{-1}+\tau_{Lh}^{-1},
\end{equation}
where where $\tau_{\textrm{phonon}}$ and $\tau_{\textrm{imp}}$ are given by \eqref{eq:tau_phonon} and \eqref{eq:tau_imp} respectively and the scattering times off the boundary are
\begin{equation}
\label{eq:tauLE}
    \tau_{Le}^{-1}=\frac{\beta}{2m^2L}\frac{\int \frac{d^2 \vec{p}}{(2\pi)^2} p^3 f^0_+(\mathbf{p})[1-f^0_+(\mathbf{p})]}{\int \frac{d^2 \vec{p}}{(2\pi)^2} f^0_+(\mathbf{p})},
\end{equation}
\begin{equation}
\label{eq:tauLH}
    \tau_{Lh}^{-1}=\frac{\beta}{2m^2L}\frac{\int \frac{d^2 \vec{p}}{(2\pi)^2} p^3 f^0_-(\mathbf{p})[1-f^0_-(\mathbf{p})]}{\int \frac{d^2 \vec{p}}{(2\pi)^2} (1-f^0_-(\mathbf{p}))}.
\end{equation}
We consider the steady state $\partial_t\vec{u}^e=\partial_t\vec{u}^h=0$ and calculate the electric current and energy current

\begin{equation}
    \vec{J}=e(n^e\vec{u}^e-n^h\vec{u}^h)
\end{equation}

\begin{equation}
    \vec{J}^E= k_BT(\Lambda^en^e\vec{u}^e+\Lambda^hn^h\vec{u}^h)
\end{equation}
where the number densities calculated from the Fermi distribution are

\begin{equation}
    n^e=\frac{N_fm}{2\pi\beta}\ln(1+e^{\beta\mu}), \qquad n^h=\frac{N_fm}{2\pi\beta}\ln(1+e^{-\beta\mu})
\end{equation}
From this, we can derive the thermoelectric coefficients in the absence of a magnetic field
\begin{equation}
    \sigma_{xx}=\frac{e^2(n^e\tau_{sh}^{-1}+n^h\tau_{se}^{-1}+(\tau_{he}^{-1}-\tau_{eh}^{-1})(n^e-n^h))}{m(\tau_{eh}^{-1}\tau_{sh}^{-1}+\tau_{he}^{-1}\tau_{se}^{-1}+\tau_{se}^{-1}\tau_{sh}^{-1})}
\end{equation}

\begin{equation}
    \Theta_{xx}=\frac{ek_B\bigg(n^e\tilde\Lambda^e-n^h\tilde\Lambda^h\bigg)}{m(\tau_{eh}^{-1}\tau_{sh}^{-1}+\tau_{he}^{-1}\tau_{se}^{-1}+\tau_{se}^{-1}\tau_{sh}^{-1})}
\end{equation}

\begin{equation}
    K_{xx}=\frac{k_B^2T\bigg(\Lambda^en^e\tilde\Lambda^e+\Lambda^hn^h\tilde\Lambda^h\bigg)}{m(\tau_{eh}^{-1}\tau_{sh}^{-1}+\tau_{he}^{-1}\tau_{se}^{-1}+\tau_{se}^{-1}\tau_{sh}^{-1})}
\end{equation}
where
\begin{equation}
    \tilde\Lambda^e=\Lambda^e(\tau_{he}^{-1}+\tau_{sh}^{-1})+\Lambda^h\tau_{eh}^{-1}
\end{equation}

\begin{equation}
    \tilde\Lambda^h=\Lambda^h(\tau_{eh}^{-1}+\tau_{se}^{-1})+\Lambda^e\tau_{he}^{-1}
\end{equation}
For momentum conservation we require
\begin{equation}
    n^e\tau_{eh}=n^h\tau_{he}
    \label{eq:mmtm_cons}
\end{equation}
We verify explicitly that the Onsager relations for the thermoelectric coefficients are satisfied if equation \eqref{eq:mmtm_cons} is satisfied.
Thus we can choose
\begin{equation}
    \tau_{eh}=\tau_0\frac{n_e+n_h}{n^h}, \qquad \tau_{he}=\tau_0\frac{n_e+n_h}{n^e}.
    \label{eq:away_from_CN}
\end{equation}
This ansatz agrees with the full numerical result obtained from \eqref{eq:Coul_drag} to within 10\% in the entire range of $\beta\mu$ and is therefore a satisfactory approximation.
By evaluating the collision integral in \eqref{eq:Coul_drag} numerically, we find

\begin{equation}
    \alpha_0\equiv \beta\tau_0^{-1}=0.15
\end{equation}

We define the dimensionless electrical conductivity as $\sigma_{ij}=\frac{N_fe^2}{2\hbar}\tilde\sigma_{ij}$. With a magnetic field and at CN, we calculate from the two fluid model that the Hall conductivity at small fields behaves like
\begin{equation}
    \lim_{B\to 0}\frac{\tilde \sigma_{xy}}{\beta\omega_c}=\frac{\beta}{m}\frac{(n^e-n^h)[(n^e+n^h)^2(\alpha_0+ \alpha_{s})-4\alpha_0^2n^en^h]}{\alpha_{s}^2(\alpha_0+\alpha_{s})^2(n^e+n^h)^2}
\end{equation}
where $\alpha_{s}=\alpha_{\textrm{imp}}+\alpha_{\textrm{ph}}$ and we have neglected the boundary scattering. We plot this quantity in Fig. \ref{fig:Sigma_xx_B} and show that the result from the two-fluid model agrees perfectly with the numerical result.

	\section{Numerical results}
	
\label{sec:num_res}
Armed with the full formalism for the QBE, we are in a position to numerically calculate the transport properties. In our companion paper \cite{Wagner} we plot the three thermoelectric coefficients as a function of $\beta\mu$. In this section we therefore focus on the behaviour of the transport coefficients at CN and in a magnetic field and we discuss the viscosity. 

\subsection{Transport at CN}

\begin{figure}
    \centering
\includegraphics[width=0.5\textwidth]{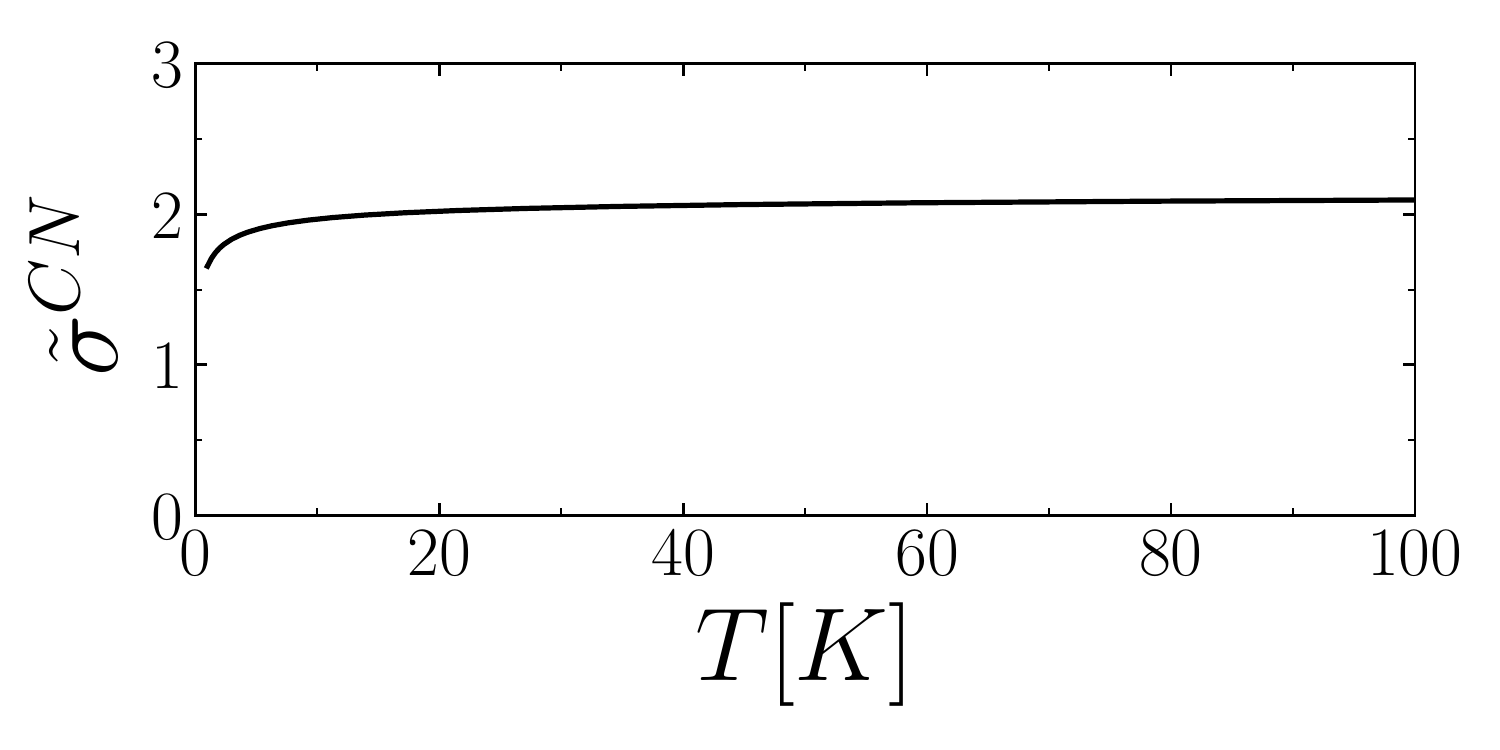}
\caption{Electrical conductivity at CN $\tilde\sigma_{xx}(\mu=0)$ plotted as a function of temperature for the canonical value $\alpha_{\textrm{ph}}=0.05$. We have also chosen $\alpha_L=\frac{\sqrt[]{\beta}}{\sqrt[]{m}L}=0.03$ (at $T=25$K) using the scale $L\sim 3\mu$m set by the sample size in \cite{Nam2017}.}
\label{fig:sigma_T}
\end{figure}

\begin{figure}
\centering
\includegraphics[width=0.5\textwidth]{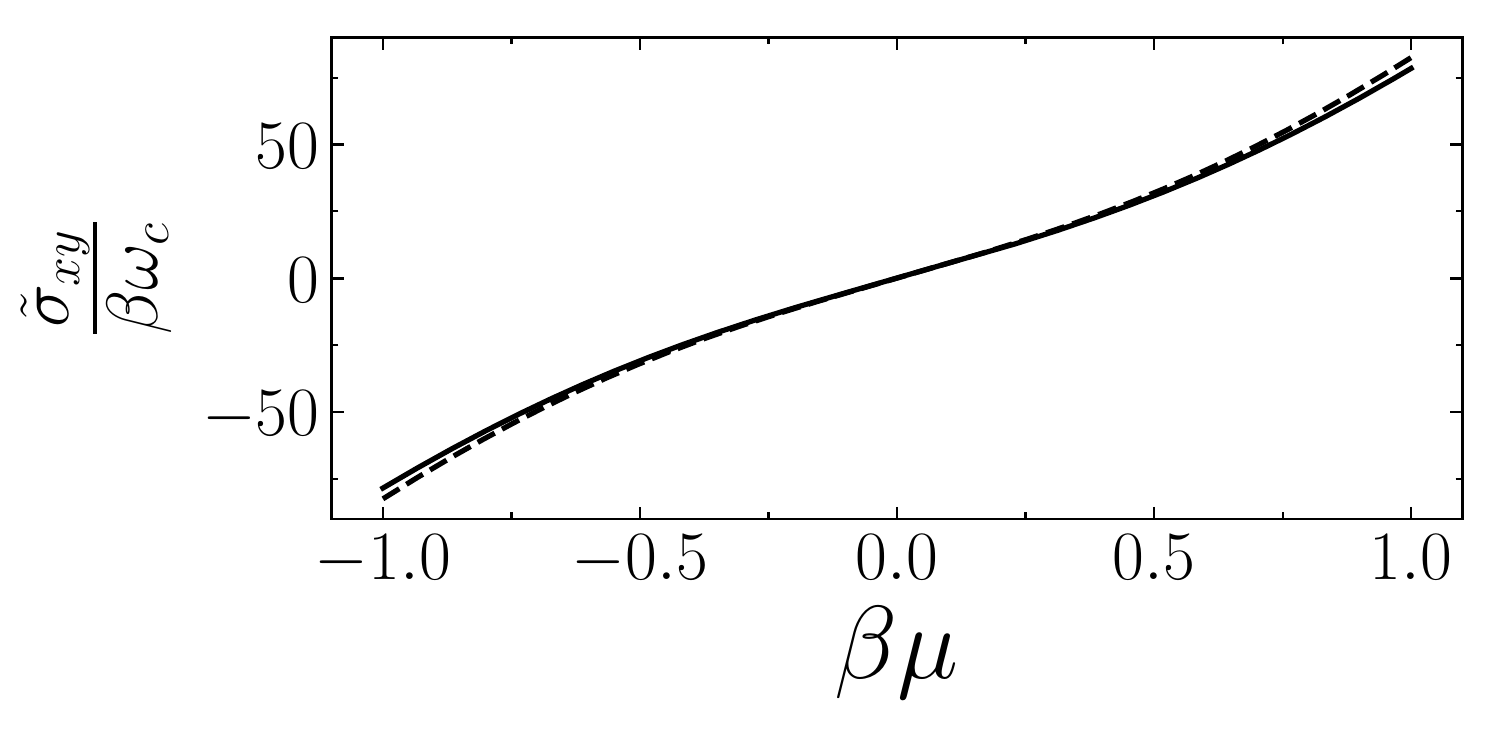}
\caption{Plot of the slope of the off-diagonal component of the electrical conductivity $\sigma_{xy}/B$ from the QBE (solid) and the two fluid model (dashed) as a function of $\beta\mu$. This plot uses the experimentally motivated value $\alpha_{\textrm{ph}}=0.05$. The two-fluid model agrees perfectly with the full QBE calculation.}
\label{fig:Sigma_xx_B}
\end{figure}

In Fig. \ref{fig:sigma_T} we show how the electrical conductivity at charge neutrality depends on temperature. In order to obtain a non-trivial temperature dependence we need to go beyond the Coulomb and phonon scattering. We assume that collisions off impurities can be neglected, as claimed in the experimental work \cite{Nam2017}. With Coulomb interactions and phonons alone, the conductivity at CN would be independent of temperature, since at CN, the conductivity would only depend on the dimensionless parameter $\alpha_\textrm{ph}$, which is temperature-independent. So the temperature-dependence is entirely due to the finite size scattering, which comes with the dimensionless number $\alpha_L$, which does depend on temperature. This figure shows qualitative agreement with the behaviour seen in Fig. 4 of \cite{Nam2017}. The thermal conductivity at charge neutrality shows the same type of behaviour as the electrical conductivity and for the same reasons.

\begin{figure}
\includegraphics[width=0.5\textwidth]{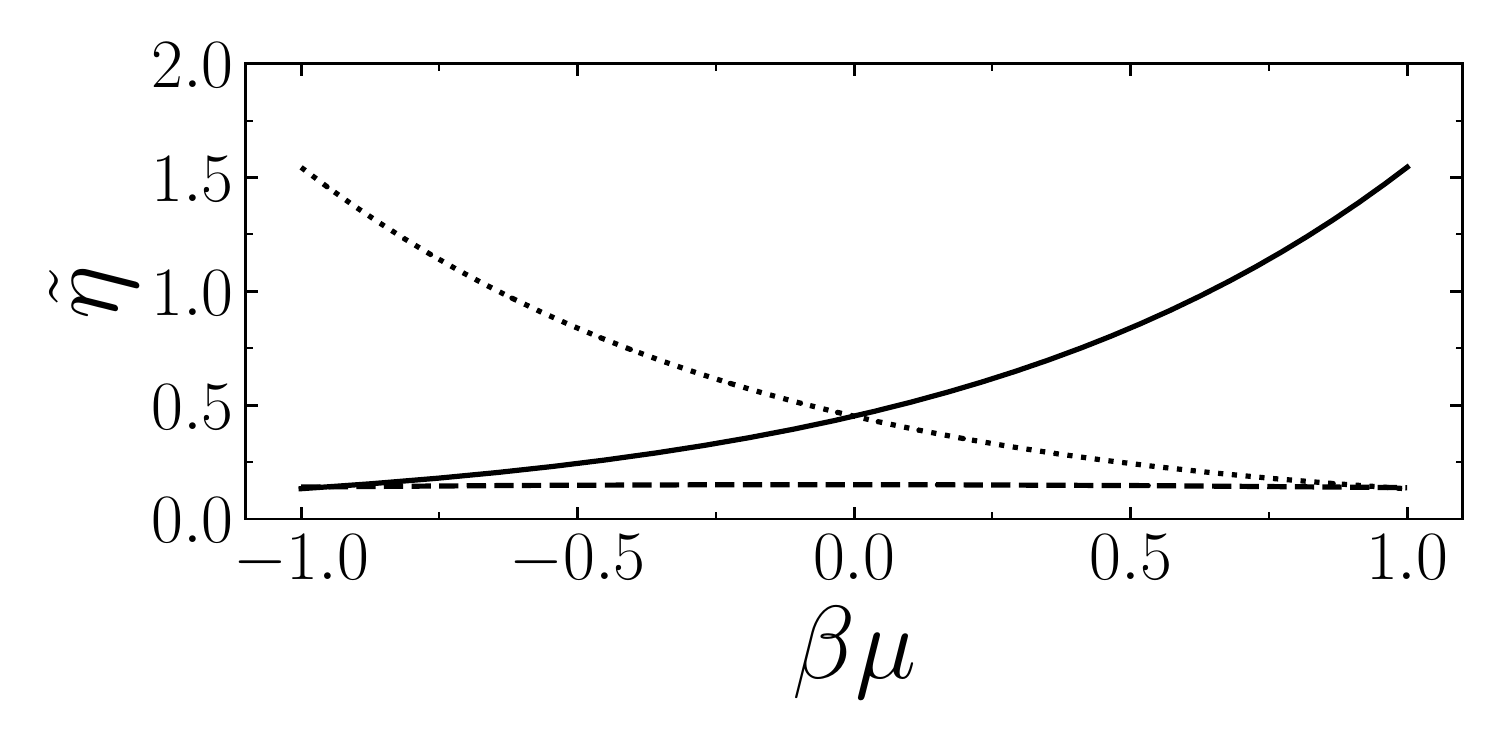}
\caption{Plot of the dimensionless viscosity $\tilde \eta_{\lambda,\lambda'}=(N_fm/\beta)^{-1}\eta_{\lambda,\lambda'}$ plotted for $\alpha_{\textrm{ph}}=0.05$. We plot $\eta_{++}$ (solid line), $\eta_{--}$ (dotted), $\eta_{+-}$ (dashed), $\eta_{-+}$ (dashed).}
\label{fig:viscosity}
\end{figure}

\subsection{Lorenz number}

\begin{figure}
\includegraphics[width=0.5\textwidth]{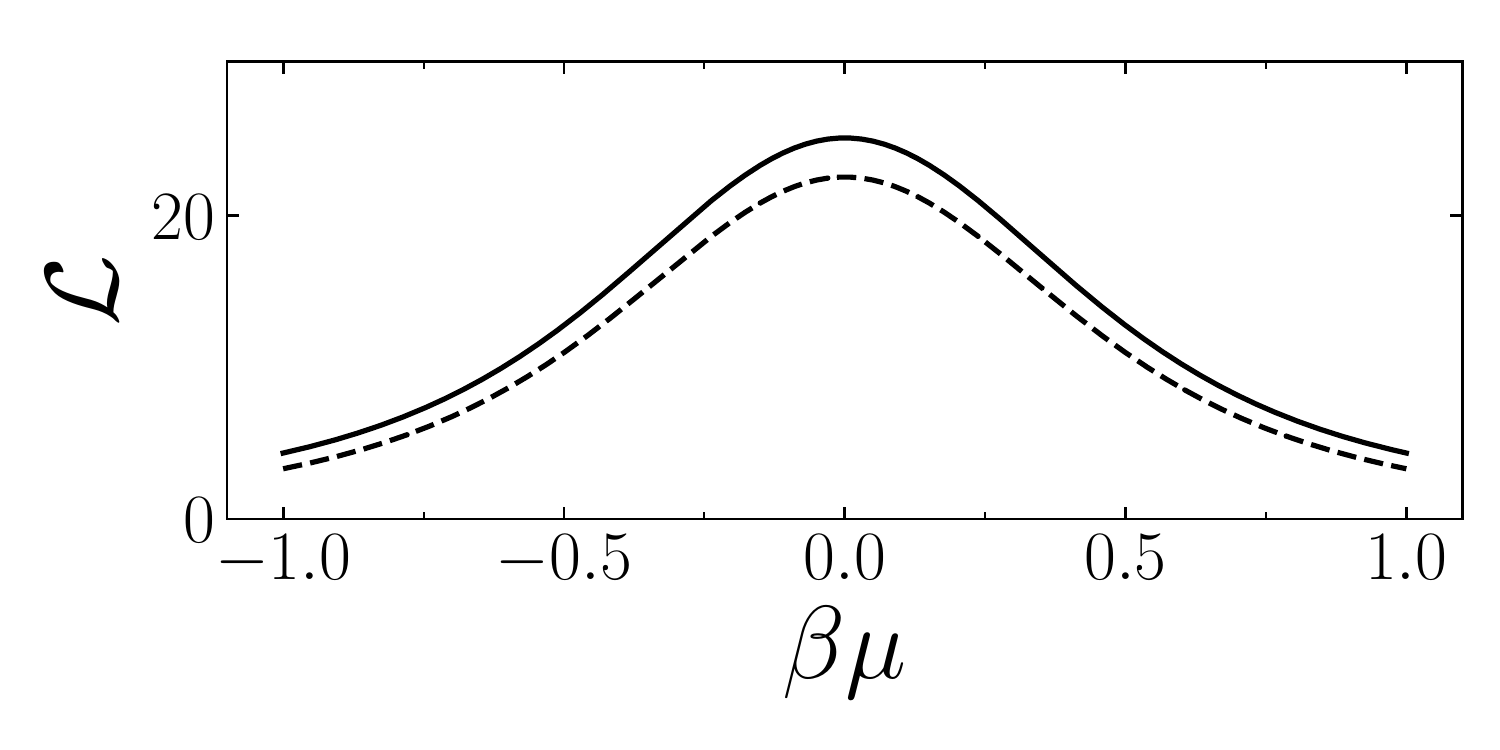}
\caption{Plot of the Lorenz number for $\alpha_\textrm{s}=\beta\tau_{s}^{-1}=0.05$. We compare the QBE results (solid) with the two-fluid model results (dashed). }
\end{figure}

From the Lorenz number $\mathcal{L}=\kappa_{xx}/\sigma_{xx}T$ and the Hall Lorenz number $\mathcal{L}_H=\kappa_{xy}/\sigma_{xy}T$ we deduce a further signature of the two-fluid model. The Lorenz number is enhanced relative to the Wiedemann-Franz (WF) law which predicts $\mathcal{L}=\pi^2/3(k_B/e)^2$. The violation of the WF law has been reported in a recent theoretical work \cite{Vignale}. On the other hand, the violation of the WF law is much less severe for the Hall Lorenz number $\mathcal{L}_H$. Both these observations can be explained in the following simple picture. We find the Lorenz number at charge neutrality 
\begin{equation}
    \mathcal{L}\equiv \frac{\kappa_{xx}}{\sigma_{xx}T}=\Lambda^2\bigg(\frac{k_B}{e}\bigg)^2\bigg(1+\frac{\tau_{eh}^{-1}+\tau_{he}^{-1}}{\tau_{s}^{-1}}\bigg)
    \label{eq:Lorenz_2fluid}
\end{equation}
where $\Lambda^e=\Lambda^h\equiv\Lambda$ at CN and $\tau_{s}^{-1}=\tau_{\textrm{imp}}^{-1}+\tau_{\textrm{phonon}}^{-1}$. We neglect scattering off the boundary in this section. From Drude theory, $\kappa_{xx}\propto \tau_{\kappa}$, where $\tau_{\kappa}^{-1}$ is the scattering rate due to all collisions that relax the energy current. At CN for an applied thermal gradient, $\vec{u}^e=\vec{u}^h$ so there is no Coulomb drag between the electrons and holes and hence only the momentum relaxing scattering limits the thermal conductivity $\tau_\kappa^{-1}=\tau_{s}^{-1}.$ On the other hand the Coulomb drag is important for the electrical conductivity, hence $\sigma_{xx}\propto \tau_\sigma$ where $\tau_\sigma^{-1}=\tau_{eh}^{-1}+\tau_{he}^{-1}+\tau_{s}^{-1}$, where we have added the scattering rates according to Matthiessen's rule \footnote{Matthiessen's rule is only valid at CN, where the electrons and holes have equal densities and scattering times.}. This immediately yields equation \eqref{eq:Lorenz_2fluid}. 

The Hall Lorenz number on the other hand is
\begin{equation}
    \mathcal{L}_H\equiv \frac{\kappa_{xy}}{\sigma_{xy}T}=\bigg(\frac{k_B}{e}\bigg)^2\Lambda^2
    \label{eq:Lorenz_H_2fluid}
\end{equation}

We can again derive this result using simple arguments. For an applied thermal gradient, electrons and holes move in the same direction, thus the only scattering along the direction of the gradient, x, is $\tau_{s}^{-1}$. Electrons and holes will be deflected in opposite directions by the applied magnetic field, so the friction in the perpendicular y direction is $\tau_{s}^{-1}+\tau_{eh}^{-1}$. This will increase the friction between the two fluids and limit the value of $\kappa_{xy}$. For an applied electric field, the electrons and holes move in opposite direction, so the friction in the x direction is $\tau_{s}^{-1}+\tau_{eh}^{-1}$. The magnetic field will deflect them in the same direction and the two fluids will feel the reduced friction $\tau_{s}^{-1}$ between them in the y direction. From these considerations, \eqref{eq:Lorenz_H_2fluid} can be shown.

Thus,
\begin{equation}
    \frac{\mathcal{L}}{\mathcal{L}_H}=1+\frac{\tau_{eh}^{-1}+\tau_{he}^{-1}}{\tau_{s}^{-1}}\gg 1
\end{equation}

From the numerics we indeed find $\mathcal{L}\approx25(k_B/e)^2$ and $\mathcal{L}_H\approx6(k_B/e)^2$.

\subsection{Viscosity}

We calculate the shear viscosity as the response of the stress tensor when a shear flow is applied to either of the particle species. 
The viscosity tensor $\eta_{\lambda\lambda'}$ is then defined via $T_\lambda^{12}=-\eta_{\lambda\lambda'}F_{\lambda'}$ and $\eta_{\lambda,\lambda'}=\frac{N_fm}{\beta}\tilde\eta_{\lambda,\lambda'}$. $\eta_{\lambda,\lambda'}$ is a measure of the friction between particle species $\lambda$ and $\lambda'$. In the numerical data FIG. \ref{fig:viscosity} we see that at large $\beta\mu$, $\eta_{++}$ dominates, since electron-electron collisions are the dominating ones. Conversely $\eta_{+-}$ decreases at large $\beta\mu$ since there are less holes present to exert a friction on the electrons. 
We also note that $\eta_{+-}\ll\eta_{++}$. This can be understood from the kinematics of collisions. Energy and momentum conservation constrain the available phase space more for electron-hole collisions than for electron-electron collisions. In addition, the matrix elements for electron-hole collisions favour large momentum exchange, which is suppressed by the potential. 
 This justifies the two-fluid model since this shows that the intra-fluid collisions of the electron and hole fluids dominate over the inter-fluid collisions and therefore we can treat the two fluids as weakly interacting.

A possible probe of the viscosity is via the negative non-local resistance \cite{Bandurin1055,Levitov2016}. In order to make this measurement quantitative, the relation between viscosity and negative nonlocal resistance must be determined which requires a full solution of the fluid equations in the relevant geometry. Another method for measuring the viscosity of graphene has been proposed using a Corbino-disc device \cite{Tomadin2014}.

We note that the famous KSS result \cite{KSS} provides a lower bound for the ratio of the shear viscosity $\eta$ to the entropy density $s$ in a strongly interacting quantum fluid, $\eta/s\geq 1/(4\pi k_B)$. In our case the entropy density is $s=n^ek_B\Lambda^e+n^hk_B\Lambda^h=(2\pi/3)mk_B/\beta$ and so
\begin{equation}
    \frac{4\pi}{k_B}\frac{\eta}{s}=24\tilde \eta\gg 1,
\end{equation}
since $\tilde\eta\gtrsim 0.5$ from Fig. \ref{fig:viscosity}. Since the bound is saturated for an infinitely strongly coupled conformal field theory, the fact that we are away from the bound is consistent with the previous arguments that we are in a weakly coupled regime and the semiclassical method is valid.

\subsection{Detailed benchmarking}

\begin{figure}
\includegraphics[width=0.5\textwidth]{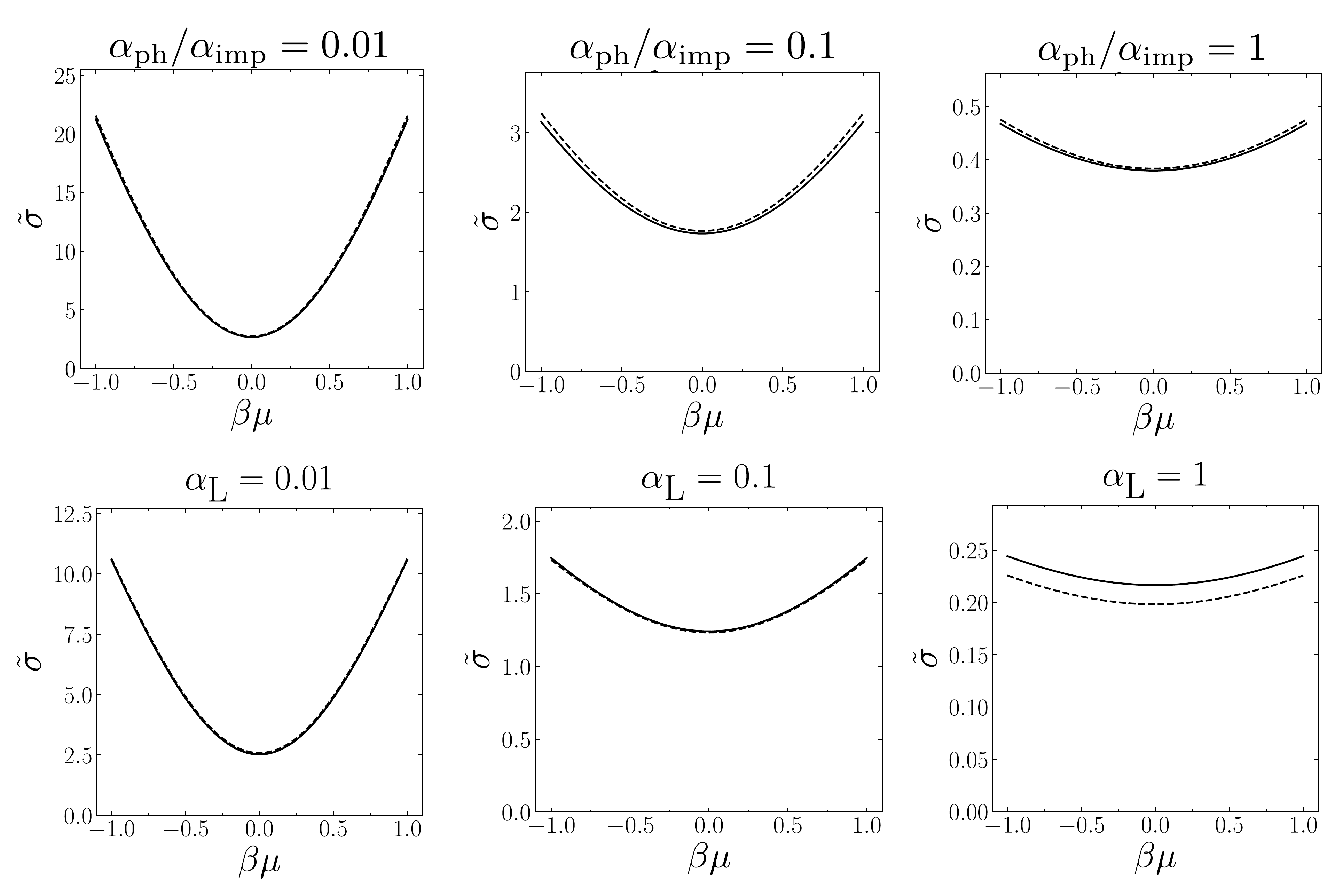}
\caption{Plots of the dimensionless conductivity defined via $\sigma_{ij}=\frac{N_fe^2}{2\hbar}\tilde\sigma\delta_{ij}$ for various values of $\alpha_L$ and $\alpha_\textrm{ph}$. We compare the QBE results (solid) with the two-fluid model results (dashed). We see that in the case of phonon or impurity scattering (top row), the results for the electrical conductivity are good for all values of the phonon coupling strength. On the other hand, for scattering off the boundary of the sample (bottom row), the agreement gets worse as the scattering off the boundary increases.}
\label{fig:sigma_benchmarking}
\end{figure}

\begin{figure}
\includegraphics[width=0.5\textwidth]{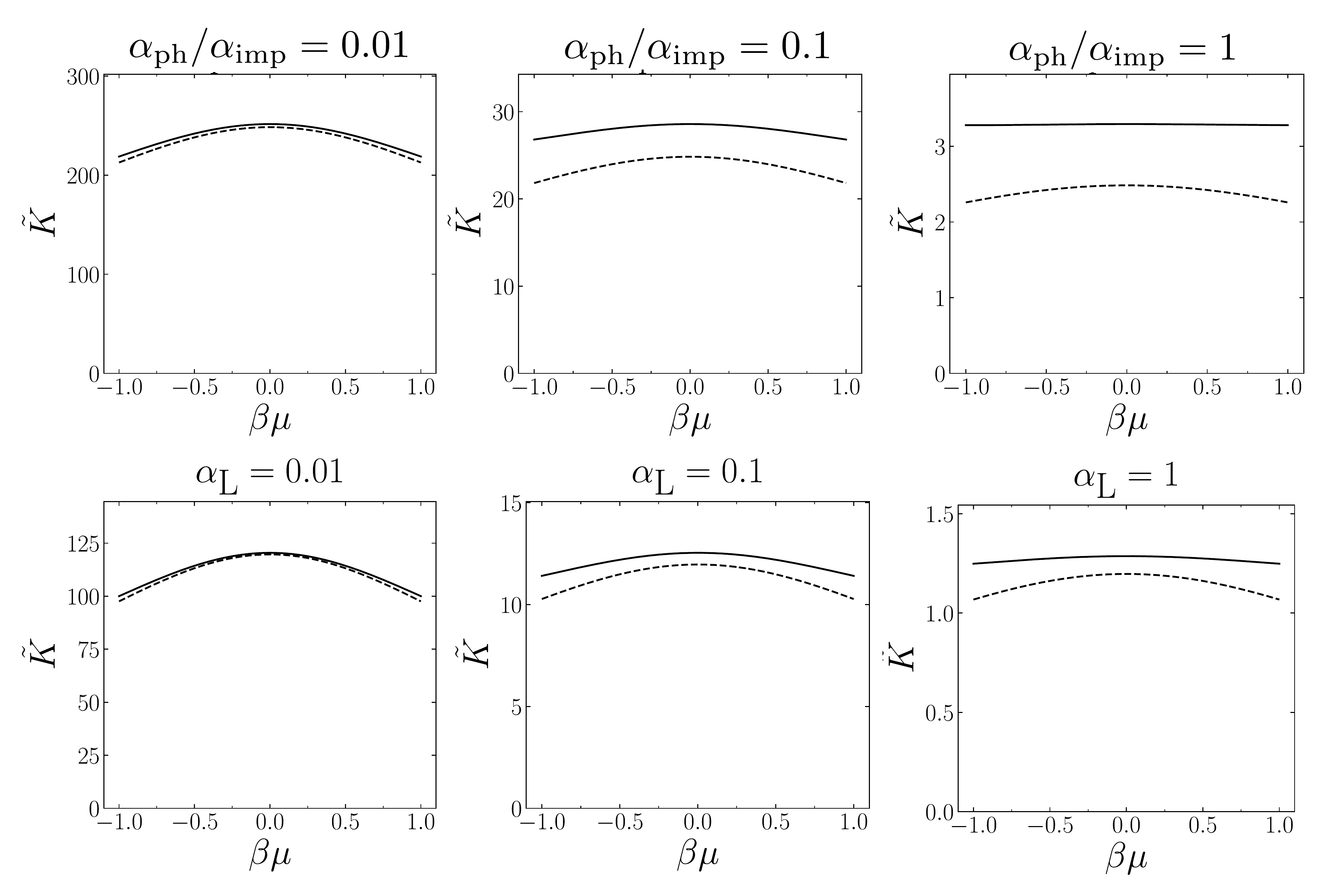}
\caption{Plots of the dimensionless thermal conductivity defined via $K_{ij}=\frac{N_fk_B^2T}{2\hbar} \tilde K\delta_{ij}$. We compare the QBE results (solid) with the two-fluid model results (dashed) for the cases of phonon or impurity scattering (top row) and for scattering off the boundary of the sample (bottom row). We see that the agreement is good for weak momentum relaxing scattering. However, as the momentum relaxing scattering is increased, the agreement gets significantly worse. Note that at large $\alpha_\textrm{ph}/\alpha_\textrm{imp}$, the QBE predicts $\tilde K$ to increase with $\beta\mu$, whereas the two fluid model predicts a decrease, so even the qualitative behaviour is incorrect in this regime.}
\label{fig:K_benchmarking}
\end{figure}

In order to assess the usefulness of the two-fluid model, we now perform a detailed analysis of the agreement between the QBE and the two-fluid model for a large range of the parameter-space of the problem. In the top rows of Figs.~\ref{fig:sigma_benchmarking} and \ref{fig:K_benchmarking} we check the agreement for phonon or impurity scattering, which is described by dimensionless strength $\alpha_\textrm{ph}$ or $\alpha_\textrm{imp}$. These two cases are identical in both the QBE and two-fluid model at fixed $\alpha$, they only differ in the temperature dependence of the dimensionless number $\alpha$. The bottom row shows the corresponding results for finite-size scattering, which is described by the dimensionless strength $\alpha_\textrm{L}$.

We see from Figs.~\ref{fig:sigma_benchmarking} and \ref{fig:K_benchmarking} that in general the agreement is very good for weak momentum-relaxing scattering, ie. small values of $\alpha_\textrm{ph}$ or $\alpha_L$. In this limit, the Coulomb-mediated electron-electron collisions are dominant and the hydrodynamic description works well. For larger values the agreement gets worse, especially in the case of the thermal conductivity. This is due to the fact that our two-fluid model only includes the equation for the first moment of the QBE. To get the thermal conductivity accurately, one would have to include the second moment as well, however this renders the two-fluid model too complicated to solve analytically, defeating the purpose of introducing it in the first place. We note however, that there is no reason to trust either the QBE or the two-fluid model in the strongly coupled regime $\alpha\gtrsim 1$. We also note the at least for the conductivity, the agreement is significantly better for the phonon and impurity scattering compared to the boundary scattering. The reason is that the boundary scattering has a scattering time that depends on momentum and in the two-fluid model we parametrize this by an average scattering time.

We note that even when the agreement with the two-fluid model fails, our QBE solution still satisfies the symmetries listed in section \ref{sec:sym}, since these are exact symmetries of the QBE. We have checked our numerical solution to find that the symmetries are indeed obeyed to an accuracy of $10^{-7}$. 

	\section{Conclusion}
	This paper sets up the Quantum Boltzmann formalism for bilayer graphene. It will serve as a reference work for numerical studies of the QBE that can be compared to experimental results. The experimentally-relevant transport quantities that we focus on are the thermo-electric coefficients (electrical conductivity, thermopower, thermal conductivity) and the shear viscosity. So far, only the electrical conductivity has been measured in experiment.  
	
	The calculation of the transport coefficients requires two ingredients: Firstly, we need to calculate the conserved currents associated with the coefficients in terms of the distribution function. In the case of the viscosity for instance, one has to calculate the stress tensor. This requires working out the coupling of BLG to a curved background metric, a calculation that is performed in this paper for the first time. Secondly, we need to work out the change in the distribution function due to the applied external fields. We use the Kadanoff-Baym equations as a starting point. The most technical part of this derivation is the calculation of the collision integral, which is performed in detail in the appendices. Once we have these ingredients, we can plug the change of the distribution function into the expressions for the conserved currents to find the linear response to the applied external fields. This allows us to read off the transport coefficients.
	
	The dominant term in the collision integral in the hydrodynamic regime of BLG will be the Coulomb interactions. However, in order to obtain a finite conductivity, we need to break the Galilean invariance of the system. There are three possible terms that can be added to the collision integral: the scattering of the electrons off (1) phonons, (2) impurities, (3) the boundary of the sample. Depending on the experimental parameters, one or the other may dominate and in this work we have calculated all three contributions.

	In the case of monolayer graphene, the electrons obey a linear dispersion relation. Energy conservation together with momentum conservation then places tight constraints on the phase space of collisions and this allows analytical results for the collision integral to be obtained \cite{Fritz2008}. A similar simplification in the case of BLG is not possible due to the quadratic energy dispersion. Due to the analogy with monolayer graphene, some previous authors have neglected scattering terms that are forbidden for monolayer graphene but allowed for BLG. We explicitly included these terms in our work. The collision integral must be evaluated numerically. 

	We derived from the QBE the two-fluid model --- a simple hydrodynamic model for the evolution of the mean fluid velocity of the electron and hole fluids. There is Coulomb drag between the two fluids and they are both subject to scattering off phonons. This model is simple enough to be able to obtain analytical formulae for the transport coefficients. We show that the two-fluid model provides quantitatively accurate results in the hydrodynamic regime where the electron-electron collisions are dominant and momentum-relaxing collisions are subdominant.
	
	Our predictions for the temperature and magnetic field dependence of the conductivity can be verified experimentally. It should be possible to add a magnetic field to the experiment and perform the measurement of the electrical and thermal conductivities. This is another interesting probe of the hydrodynamic regime in BLG and can be used to check the agreement between the experimental behaviour and the theoretical predictions. 
	
	Our formalism can be adapted to consider BLG far from charge neutrality by modifying the screening calculation. Due to the quadratic dispersion, we expect that in the $\beta\mu\gg1$ regime, one recovers the standard Fermi liquid results.
	
	It is also possible to generalize the formalism to treat multilayer graphene.   A further possible avenue of research is to extend the present formalism to finite frequencies. Besides adding an extra term to the collision integral corresponding to the time-derivative in the Boltzmann equation, this may require taking into account the off-diagonal components of the probability distribution matrix as well as considering the \textit{Zitterbewegung} contributions. 
	
	Finally, another direction of research is the calculation of the Hall viscosity, which has been measured experimentally in monolayer graphene and BLG experiments \cite{berdyugin2018measuring}.
	
	\section{Acknowledgements}
	We would like to acknowledge helpful discussions with Philipp Dumitrescu. We thank Christopher Herzog and Dam Thanh Son for their comments on the KSS bound. This work was supported by EP/N01930X/1 and EP/S020527/1. Statement  of  compliance  with  EPSRC  policy  framework  on  research  data:  This  publication is theoretical work that does not require supporting research data.
	
	\emph{Note added:} During the completion of this work we became aware of related work \cite{Vignale} which looks at the thermoelectric properties for BLG. However, in our work, we present a detailed derivation of the quantum Boltzmann formalism and we have a different form of the collision integral. In addition, we deduce the effect of the experimentally relevant finite-size effect as well as phonon scattering. We aim to provide quantitative results that can be compared to current and future experimental data.

	
	\appendix

	\section{Detailed calculation for Coulomb screening in a homogeneous metric}
	\label{sec:CoulbCur}
	In this subsection, we will show how to modify the calculation for screening momentum in the homogeneous metric
	\begin{equation}
	g_{ij}=\delta_{ij}+\delta g_{ij},
	\end{equation}
	where $\delta g_{ij}$ is space-time independent. The effective Hamiltonian is given by 
	\begin{multline}
	\label{eq:curvHam1}
	\cH=\sum_{\lambda f}\int d^2 x\sqrt{g} \hat{\Psi}^\dagger_{\lambda f}(\mx)\bigg(-\lambda\frac{g^{ij}\partial_i \partial_j}{2m}-\mu\bigg)\hat{\Psi}_{\lambda f}(\mx)
	\\+\frac{1}{2}\sum_{\{\lambda_i\},f,f'}\int d^2 x\sqrt{g} \int d^2 y\sqrt{g}\hat{\Psi}^\dagger_{\lambda_1 f}(\mx)\hat{\Psi}_{\lambda_2 f}(\mx)\\\times\frac{2\pi \alpha}{|\mx-\my|}\hat{\Psi}^\dagger_{\lambda_3 f'} (\my) \hat{\Psi}_{\lambda_4 f'}(\my),
	\end{multline}
	where here we define 
	\begin{equation}
	|\mx -\my|=\sqrt{g_{ij}(x^i-y^i)(x^j-y^j)}.
	\end{equation}
	Note that this definition only works for a homogeneous metric, for a general metric, we need to replace $|\mx -\my|$ by the geodesic distance. We follow \cite{Fujikawa1981} and define the rescaled field as 
	\begin{equation}
	\tilde{\hat{\Psi}}(\mx)=g^{1/4}\hat{\Psi}(\mx), \qquad \tilde{\hat{\Psi}}^\dagger(\mx)=g^{1/4}\hat{\Psi}^\dagger(\mx).
	\end{equation}
	We can rewrite the Hamiltonian in the momentum space as 
	\begin{multline}
	\label{eq:Hamcurk}
	\cH=\sum_{\lambda f}\int \frac{d^2 k}{(2\pi)^2} \tilde{\hat{\Psi}}^\dagger_{\lambda f}(\mk)\bigg(\lambda\frac{g^{ij}k_i k_j}{2m}-\mu\bigg)\tilde{\hat{\Psi}}_{\lambda f}(\mk)
	\\+\frac{1}{2\sqrt{g}}\sum_{\{\lambda_i\},f,f'}\int \frac{d^2 k}{(2\pi)^2} \frac{d^2 k'}{(2\pi)^2} \frac{d^2 q}{(2\pi)^2}\tilde{\hat{\Psi}}^\dagger_{\lambda_1 f}(\mk-\mq)\tilde{\hat{\Psi}}_{\lambda_2 f}(\mk)\\\times\frac{2\pi \alpha}{|\mq|} \tilde{\hat{\Psi}}^\dagger_{\lambda_3 f'} (\mk'+\mq)\tilde{\hat{\Psi}}_{\lambda_4 f'}(\mk').
	\end{multline}
	where the Fourier transformation is given as follows
	\begin{equation}
	f(\mk)=\int d^2 x e^{i  k_i x^i} f(\mx), \qquad f(\mx)=\int \frac{d^2 k}{(2\pi)^2}e^{-i k_i x^i} f(\mk)
	\end{equation}
	and we define 
	\begin{equation}
	\label{eq:norm}
	|\mq|=\sqrt{g^{ij}q_i q_j}=\sqrt{g_{ij}q^i q^j}.
	\end{equation}
	The screened Coulomb interaction is given by 
	\begin{align}
	\label{eq:CoulombCurp}
	\tilde{\cV}(\mq,0)=\frac{\tilde{V}(|\mq|)}{1+\tilde{\varPi}^0(\mq,0) \tilde{V}(|\mq|)},
	\end{align}
	where $V(|\mq|)$ can be read off from \eqref{eq:Hamcurk} 
	\begin{equation}
	\tilde{V}(|\mq|)=\frac{1}{\sqrt{g}}\frac{2\pi \alpha}{|\mq|}.
	\end{equation}
	From the Hamiltonian \eqref{eq:Hamcurk}, we see that the propagator of $\tilde{\hat{\Psi}}$ is given by
	\begin{equation}
	\label{eq:MGreencur}
	\tilde{G}^{\lambda\lambda'}(i\omega_n,\mathbf{k})=\frac{1}{i\omega_n -\lambda \frac{|\mk|^2}{2m}+\mu}\delta_{\lambda \lambda'}.
	\end{equation}
	We can repeat the calculation in section \ref{sec:Coul} to obtain the susceptibility 
	\begin{multline}
	\label{eq:suscur}
	\tilde{\varPi}^0(\mathbf{q},0)=N_f\sum_{\lambda,\lambda'}\int{\frac{d^2 \mathbf{k}}{(2\pi)^2}}\frac{f^0_{\lambda'}(|\mathbf{k+q}|,\mu)-f^0_{\lambda}(|\mk|,\mu)}{\lambda'\frac{|\mathbf{k}+\mathbf{q}|^2}{2m}-\lambda \frac{|\mk|^2}{2m}}\\\times F(\lambda,\lambda',\mk,\mk+\mq).
	\end{multline}
	From the definition \eqref{eq:norm}, we introduce the new variable
	\begin{equation}
	\label{eq:transk}
	K_a=e^i_a k_i, \qquad k_i = e^a_i K_a, 
	\end{equation}
	and obtain 
	\begin{equation}
	\label{eq:transk1}
	k_i p^i=g^{ij}k_i p_j =  e^i_a e^j_a k_i p_j =K_a P_a.
	\end{equation}
	By the means of transformation \eqref{eq:transk}, we have 
	\begin{equation}
	\label{eq:transk2}
	d^2k =\det(e^a_i)d^2 K =\sqrt{g} d^2 K.
	\end{equation}
	Combining equations \eqref{eq:transk1}, \eqref{eq:transk2}, and \eqref{eq:suscur} we arrive at 
	
	\begin{equation}
	\tilde{\varPi}^0(\mathbf{q},0)=\sqrt{g}\varPi^0(|\mathbf{q}|,0).
	\end{equation} 
	Thus we have 
	\begin{align}
	\tilde{\cV}(\mq,0)=\frac{2\pi \alpha}{\sqrt{g}}\frac{1}{|\mq|+q_{TF}(|\mq|)}.
	\end{align}
	Note that the calculation in this subsection only valid for a homogeneous metric. The screening potential for general metric should be calculated in a completely different manner.
	\section{Detailed derivation of energy current and stress tensor operators}
	\label{sec:operaotsApp}
	In this Appendix, we will present the detailed derivation of equation \eqref{eq:energykin} and \eqref{eq:stresskin}. he exact low energy wave function can be calculated directly by diagonalizing the Hamiltonian \eqref{eq:Ham}, we obtain
	\begin{align}
	\label{eq:WaveK1e}
	\psi_{cK}^\sigma=\frac{|\mathbf{k}| v_F/\bar k}{\sqrt{2}\gamma_1}\begin{pmatrix}
	-\frac{\bar k}{k}\frac{|\mathbf{k}|^2 v_F^2+\gamma_1^2}{\gamma_1 v_F} \\ \frac{|\mathbf{k}|^2 v_F^2+\gamma_1^2}{  \gamma_1 v_F} \\\bar{k}\\-\bar{k}
	\end{pmatrix}\otimes|\sigma\rangle,
	\end{align}
	\begin{align}
	\label{eq:WaveK2e}
	\psi_{vK}^\sigma=\frac{|\mathbf{k}| v_F/\bar{k}}{\sqrt{2}\gamma_1}\begin{pmatrix}
	\frac{\bar k}{k}\frac{|\mathbf{k}|^2 v_F^2+\gamma_1^2}{\gamma_1 v_F} \\ \frac{|\mathbf{k}|^2 v_F^2+\gamma_1^2}{\gamma_1 v_F} \\-\bar{k}\\-\bar{k}
	\end{pmatrix}\otimes|\sigma\rangle,
	\end{align}
	where $k=k_x+ik_y$ and $\bar{k}=k_x-ik_y$. We see that when we take the approximation $v_F |\mathbf{k}| \ll \gamma_1$, we recover the low energy results up to a gauge. 
	\subsection{Derivation of  \eqref{eq:energykin}}
	In order to calculate the commutator in \eqref{eq:Energycurrent}, we need to write down the explicit form of the Hamiltonian in second quantization language
	\begin{equation}
	\label{eq:htot}
	\cH=\sum_{\xi}\int \frac{d^2 k}{(2\pi)^2}\hat{\Psi}^\dagger_\xi(\mk) \hat{H}_{\xi}(\mk) \hat{\Psi}_\xi(\mk)+\hat{V}_C,
	\end{equation}
	with $\hat{H}_{\xi}(\mk)=\hat{\mathbf{H}}_{\xi}(\mk,0)$ and 
	\begin{multline}
	\hat{V}_C=\frac{1}{2}\sum_{f f'a b}\int \frac{d^2 k}{(2\pi)^2}\frac{d^2 k'}{(2\pi)^2}\frac{d^2 q}{(2\pi)^2}c^\dagger_{f a}(\mk)c_{f a}(\mk-\mq)\\\times V_C(\mq)c^\dagger_{f' b}(\mk')c_{f'b }(\mk'+\mq).
	\end{multline}
	Combining equations \eqref{eq:hkin} and \eqref{eq:htot} and using the anti-commutation relation 
	\begin{equation}
	\{c^\dagger_{i}(\mk),c_j(\mk')\}=\delta_{ij}\delta(\mk-\mk'),
	\end{equation}
	we have 
	\begin{multline}
	\label{eq:Energy1}
	[\mathbf{H}_{\xi}(\mq),\cH]=\int \frac{d^2 k}{(2\pi)^2}\hat{\Psi}^\dagger_\xi(\mk-\mq)\left(\hat{\mathbf{H}}_{\xi}(\mk,\mq)\hat{H}_{\xi}(\mk)\right.\\\left.-\hat{H}_{\xi}(\mk-\mq)\hat{\mathbf{H}}_{\xi}(\mk,\mq)\right)\hat{\Psi}_\xi(\mk)+[\mathbf{H}_{\xi}(\mq),\hat{V}_C].
	\end{multline}
	We now will calculate the last contribution. We can write down explicitly the commutator as 
	\begin{multline}
	[\mathbf{H}_{\xi}(\mq),\hat{V}_C]=\sum_s\int \frac{d^2 k}{(2\pi)^2}\frac{d^2 k'}{(2\pi)^2}\frac{d^2 q'}{(2\pi)^2} \hat{h}_{\xi a b}(\mk,\mq)\\\times\left(c^\dagger_{\xi s a}(\mk-\mq)c_{\xi s b }(\mk-\mq')-c^\dagger_{\xi s a}(\mk-\mq+\mq')c_{\xi s b }(\mk)\right)\\
	\times V_C(\mq')\sum_{f'c}c^\dagger_{f' c }(\mk')c_{f' c }(\mk'-\mq').
	\end{multline}
	Up to linear order in perturbation, and due to the fact that $\hat{h}_{\xi a a}=0$ , the only nonzero contribution comes from $\mathbf{q}'=0$. We can rewrite the above equation as 
	\begin{multline}
	[\mathbf{H}_{\xi}(\mq),\hat{V}_C]=\sum_s\int \frac{d^2 k}{(2\pi^2)} \hat{\mathbf{H}}_{\xi a b}(\mk,\mq)\\\times\left(c^\dagger_{\xi s a}(\mk-\mq)c_{\xi s b }(\mk)-c^\dagger_{\xi s a}(\mk-\mq)c_{\xi s b }(\mk)\right) V_C(0)N_0=0,
	\end{multline}
	where $N_0$ is the background charge. The contribution of the kinetic part to the energy current comes from the first term of equation \eqref{eq:Energy1}.
	We can read off the spatially independent current density by taking the $\mathbf{q}\rightarrow 0$ limit of equation \eqref{eq:Energycurrent}
	\begin{multline}
	\mathbf{J}_{kin}^E(\mq=0)=\sum_{\xi}\mathbf{J}^E_\xi(\mq=0)\\=\sum_{\xi}\int \frac{d^2 k}{(2\pi)^2}\hat{\Psi}^\dagger_\xi(\mk)\hat{\mathbf{J}}^E_{\xi}(\mk,\mq=0)\hat{\Psi}_\xi(\mk),
	\end{multline}
	where 
	\begin{equation}
	\hat{J}^{E;1}_{\xi}(\mk,\mq=0)=\begin{pmatrix}
	k_1 v_F^2 &0 &\xi \frac{v_F \gamma_1}{2}  &0\\0 &k_1 v_F^2 &0 &\xi \frac{v_F \gamma_1}{2}\\\xi \frac{v_F \gamma_1}{2} &0 &k_1 v_F^2 &0\\0 &\xi \frac{v_F \gamma_1}{2} &0 &k_1 v_F^2
	\end{pmatrix},
	\end{equation}
	\begin{equation}
	\hat{J}^{E;2}_{\xi}(\mk,\mq=0)=\begin{pmatrix}
	k_2 v_F^2 &0 &-i\xi \frac{v_F \gamma_1}{2}  &0\\0 &k_2 v_F^2 &0 &i\xi \frac{v_F \gamma_1}{2}\\i\xi \frac{v_F \gamma_1}{2} &0 &k_2 v_F^2 &0\\0 &-i\xi \frac{v_F \gamma_1}{2} &0 &k_2 v_F^2
	\end{pmatrix}.
	\end{equation}
	The contribution from low energy bands can be calculated explicitly by substituting wave functions \eqref{eq:WaveK1e}, \eqref{eq:WaveK2e} into the above equation and then making the approximation $v_F |\mathbf{k}| \ll \gamma_1$. We can obtain the kinetic part of the energy current by adding the contribution from $K$ and $K'$ valleys to derive \eqref{eq:energykin}.

	\subsection{Derivation of \eqref{eq:stresskin}}
	In this subsection, we will present the detailed derivation of \eqref{eq:stresskin}. We rewrite the definition \eqref{eq:Tijdef} in terms of the vierbein instead of the metric as follows. First, we introduce the vierbein $e^i_a$ with following definition
	\begin{equation}
	\label{eq:vie}
	e^i_a(\x) e^j_b(\x) \delta^{ab}=g^{ij}(\x).
	\end{equation}
	Using the definition \eqref{eq:Tijdef} and the symmetry of the stress tensor operator $T^{ij}$, we have the new definition of $T^{ij}$
	\begin{multline}
	\label{eq:Tijdef1}
	\!\!\!\!\!T^{ij}(\x)=\\\frac{1}{2}\left(e^i_a(\x)\frac{\delta \cS(e^{k}_c,\tilde{\hat{\Psi}}^\dagger,\tilde{\hat{\Psi}})}{\delta e_{aj}(\x)}+e^j_b(\x)\frac{\delta \cS(e^{k}_c,\tilde{\hat{\Psi}}^\dagger,\tilde{\hat{\Psi}})}{\delta e_{bi}(\x)}\right)\arrowvert_{e^{k}_c=\delta^k_c}.
	\end{multline}
	The zero momentum component of the stress tensor $T^{ij}(\mathbf{q}=0)$ is given by the response of the system to a homogeneous perturbation of the local metric $\delta g_{ij}(\x)=\delta g_{ij}$. From the definition \eqref{eq:Tijdef}, the only contribution to the stress tensor $T^{ij}(\mathbf{q}=0)$ comes from the Hamiltonian \footnote{We see that the first term of the effective action $i\tilde{\hat{\Psi}}_{\xi}^\dagger \partial_t \tilde{\hat{\Psi}}_{\xi}$ has no variation with respect to the homogeneous perturbation of the metric.}. The coupling of the kinetic part of the Hamiltonian with the spatially uniform vierbein is given by \footnote{Since we only consider a spatially independent perturbation of the metric, the spin connection vanishes.}
	\begin{equation}
	\label{eq:Tijkin}
	H^{kin}_{\xi}=\int \frac{d^2 k}{(2\pi)^2} \tilde{\hat{\Psi}}_{\xi}^{\dagger}(\mathbf{k})\tilde{\cH}_{\xi}(\mathbf{k})\tilde{\hat{\Psi}}_{\xi}(\mathbf{k}),
	\end{equation}
	with
	\begin{equation}
	\tilde{\cH}_{\xi}(\mathbf{k})=\xi v_F\left(
	\begin{array}{c|c}
	\begin{matrix} 0 & 0\\0&0\end{matrix}&  e_i^a\sigma_a k^i\\
	\hline
	e_i^a\sigma_a k^i&\begin{matrix} 0 & 0\\0&0\end{matrix}
	\end{array}
	\right).
	\end{equation}
	Combining \eqref{eq:Tijdef1} and \eqref{eq:Tijkin}, one can derive the kinetic contribution to the spatially independent part of the stress tensor operator from each valley
	\begin{multline}
	T^{ij}_\xi(\mq=0)=\xi \frac{v_F}{2}\\\times\int \frac{d^2 k}{(2\pi)^2} \hat{\Psi}^\dagger_{\xi}(\mk)  \left(
	\begin{array}{c|c}
	\begin{matrix} 0 & 0\\0&0\end{matrix}& (\sigma^i k^j+\sigma^j k^i)\\
	\hline
	(\sigma^i k^j+\sigma^j k^i)&\begin{matrix} 0 & 0\\0&0\end{matrix}
	\end{array}
	\right) \hat{\Psi}_\xi(\mk).
	\end{multline} 
	We plug in the explicit form of wave functions \eqref{eq:WaveK1e} and \eqref{eq:WaveK2e} of the low energy band and use the approximation $v_F |\mathbf{k}| \ll \gamma_1$ in the above equation and obtain \eqref{eq:stresskin}.
	\subsection{Derivation of \eqref{eq:TC}}
	In this subsection, we will present the detailed derivation for the interaction contribution to the stress tensor. We rewrite the Coulomb interaction \eqref{eq:Coulombcurved} 
    \begin{equation}
	\label{eq:Coulombcurved1}
	\hat{V}_C=\frac{1}{2}\int \frac{d^2 p}{(2\pi)^2} \frac{2\pi \alpha}{\sqrt{g}(|\mathbf{p}|+q_{TF})}\tilde{\rho}^{\text{eff}}(-\mathbf{p})\tilde{\rho}^{\text{eff}}(\mathbf{p}),
	\end{equation}
	where in $\tilde{\rho}^{\text{eff}}(\mathbf{p})$, we replace the field operator $c_{\lambda f}(\mk)$ by the rescaled one with homogeneous metric perturbation 
	\begin{equation}
	    \tilde{c}_{\lambda f}(\mk)=g^{1/4}c_{\lambda f}(\mk).
	\end{equation}
	The metric dependence of $\hat{V}_C$ in the linear transport formalism is in the screened Coulomb potential \begin{equation}
	\mathcal{V}_C(\mathbf{p})=\frac{2\pi \alpha}{\sqrt{g}(|\mathbf{p}|+q_{TF})}.    
	\end{equation}
	We take the derivative of $\mathcal{V}_C(\mathbf{p})$ with respect to $g_{ij}$ and obtain
	\begin{equation}
	\label{eq:derivativeVc}
	\frac{\delta \mathcal{V}_C(\mathbf{p})}{\delta g_{ij}}\arrowvert_{g_{ij}=\delta_{ij}}=2\pi\alpha\Big[-\frac{p^i p^j}{p}\frac{1}{(p+q_{TF})^2}+\delta^{ij}\frac{1}{p+q_{TF}}\Big].    
	\end{equation}
	We then plug \eqref{eq:derivativeVc} and \eqref{eq:Coulombcurved1} into the definition \eqref{eq:Tijdef} with $\delta g_{ij}(\mathbf{x})= \delta g_{ij}$ to obtain \eqref{eq:TC}.
	
	\section{Density matrix formalism for BLG}
	\label{sec:BoltzA}
	\subsection{Generalizing the quantum Boltzmann equation to charge conductivity}	
	In this section, we will derive the density matrix formalism for BLG. The derivation follows closely Ref \cite{Fritz2008} and the classic book \cite{Kadanoff}. The effective Hamiltonian for each flavor is given by 
	\begin{equation}
	\label{eq:Hameff10}
	H^f_{0ab}=-\frac{1}{2m}\begin{pmatrix} 
	0  &(\pi^\dagger)^2 \\
	\pi^2 & 0 
	\end{pmatrix}, \qquad \Psi=\begin{pmatrix} 
	\psi^f_{1}(\mathbf{x})   \\
	\psi^f_{2}(\mathbf{x})
	\end{pmatrix}.
	\end{equation}
	We begin with the modification of equations (8.27) and (8.28) of Ref. \cite{Kadanoff}. We define the density matrix for each flavor as 
	\begin{multline}
	\label{eq:GrL}
	G^{<}_{fa fb}(1,1',U)=\\i\begin{pmatrix} 
	\<\psi^{f\dagger}_{1}(\mx_{1'},t_{1'})\psi^f_{1}(\mx_{1},t_{1})\>_U  &\<\psi^{f\dagger}_{2}(\mx_{1'},t_{1'})\psi^f_{1}(\mx_{1},t_{1})\>_U \\
	\<\psi^{f\dagger}_{1}(\mx_{1'},t_{1'})\psi^f_{2}(\mx_{1},t_{1})\>_U & \<\psi^{f\dagger}_{2}(\mx_{1'},t_{1'})\psi^f_{2}(\mx_{1},t_{1})\>_U 
	\end{pmatrix},
	\end{multline} 
	\begin{multline}
	G^>_{fa fb}(1,1',U)=\\-i\begin{pmatrix} 
	\label{eq:GrR}
	\<\psi^f_{1}(\mx_{1},t_{1})\psi^{f\dagger}_{1}(\mx_{1'},t_{1'})\>_U  &\<\psi^f_{1}(\mx_1,t_1)\psi^{f\dagger}_{2}(\mx_{1'},t_{1'})\>_U \\
	\<\psi^f_{2}(\mx_1,t_1)\psi^{f\dagger}_{1}(\mx_{1'},t_{1'})\>_U & \<\psi^f_{2}(\mx_1,t_1)\psi^{f\dagger}_{2}(\mx_{1'},t_{1'})\>_U 
	\end{pmatrix},
	\end{multline}
	where $a,b=1,2$ are the sub-lattice indices, $f=1,..,4$ is the flavor index. 	
	We define the perturbed expectation value as 
	\begin{equation}
	\<\cO_1(t_1)\cO_2(t_2)\>_U=\<\mathbf{U}^\dagger(t_1)\cO_1(t_1)\mathbf{U}(t_1)\mathbf{U}^\dagger(t_2)\cO_2(t_2)\mathbf{U}(t_2)\>,
	\end{equation}
	and we define the unitary transformation as 
	\begin{equation}
	\mathbf{U}(t)=T\left\{exp\left[-i\int_{-\infty}^{t}d2 U(2)\rho(2)\right] \right\} 
	\end{equation}
	where $U(\mx,t)$ is the applied scalar potential and $\rho(\mx,t)$ is the density operator. By construction, we notice that 
	\begin{equation}
	G^>_{fa f'b}(1,1',U)=G^<_{fa f'b}(1,1',U)=0, \qquad (f\neq f')
	\end{equation}
	since the perturbation and interactions conserve the flavor.
	
	The Green's functions $G^{<,>}_{ab}(1,1',U)$ satisfies the following, so-called Kadanoff-Baym equations of motion
	\begin{widetext}
	\begin{multline}
	\label{eq:EOM1}
	\left[i\frac{\partial}{\partial t_{1}}-U(1)\right]G^{<,>}(1,1^{'};U)-G^{<,>}(1,1^{'};U)H_{0}(1)=\int_{-\infty}^{t_1} d \bar{1} \left[\Sigma^{>}(1,\bar{1};U)-\Sigma^{<}(1,\bar{1};U)\right]G^{<,>}(\bar{1},1';U)\\
	-\int_{-\infty}^{t_{1'}}d \bar{1}\Sigma^{<,>}(1,\bar{1};U)\left[G^{>}(\bar{1},1';U)-G^{<}(\bar{1},1';U)\right],
	\end{multline}
	\begin{multline}
	\label{eq:EOM2}
	\left[-i\frac{\partial}{\partial t_{1'}}-U(1')\right]G^{<,>}(1,1^{'};U)-H_{0}(1')G^{<,>}(1,1^{'};U)=\int_{-\infty}^{t_1} d \bar{1} \left[G^{>}(1,\bar{1};U)-G^{<}(1,\bar{1};U)\right]\Sigma^{<,>}(\bar{1},1';U)\\
	-\int_{-\infty}^{t_{1'}}d \bar{1}G^{<,>}(1,\bar{1};U)\left[\Sigma^{>}(\bar{1},1';U)-\Sigma^{<}(\bar{1},1';U)\right],
	\end{multline}
	in which we omit the indices, the equations \eqref{eq:EOM1} and \eqref{eq:EOM2} need to be considered as  matrix equations. 
	The self-energy matrix in Born collision approximation is given by 
	\begin{multline}
	\label{eq:selfE1}
	\Sigma^{>,<}_{\alpha \beta}(1,1';U)=(-i^2)\int d\mx_2 d\mx_{2'}V_C(\mx_1-\mx_2)V_C(\mx_{1'}-\mx_{2'})\Big[G^{>,<}_{\alpha\beta}(1,1';U)G^{>,<}_{\gamma\delta}(2,2';U)G^{<,>}_{\delta\gamma}(2',2;U)\\-G^{>,<}_{\alpha\gamma}(1,2';U)G^{<,>}_{\gamma\delta}(2',2;U)G^{>,<}_{\delta \beta}(2,1';U)\Big]_{t_2=t_1,t_{2'}=t_{1'}}.
	\end{multline}
	Equation \eqref{eq:selfE1} is a matrix equation, both left and right sides are $8\times 8$ matrix.
	
	At this point we are ready to derive the equation of motion for the density matrix. We subtract \eqref{eq:EOM2} from \eqref{eq:EOM1} to obtain
	\begin{multline}
	\label{eq:Kinetic1}
	\Big[i\left(\frac{\p}{\p_{t_1}}+\frac{\p}{\p_{t_{1'}}}\right)-U(1)+U(1')\Big]G^{<}(1,1';U)-G^{<}(1,1';U)H_{0}(1)+H_0(1')G^{<}(1,1^{'};U)=\\\int_{-\infty}^{t_1} d \bar{1} \left[\Sigma^{>}(1,\bar{1};U)-\Sigma^{<}(1,\bar{1};U)\right]G^{<}(\bar{1},1';U)
	+\int_{-\infty}^{t_{1'}}d \bar{1}G^{<}(1,\bar{1};U)\left[\Sigma^{>}(\bar{1},1';U)-\Sigma^{<}(\bar{1},1';U)\right]\\
	-\int_{-\infty}^{t_1} d \bar{1} \left[G^{>}(1,\bar{1};U)-G^{<}(1,\bar{1};U)\right]\Sigma^{<}(\bar{1},1';U)
	-\int_{-\infty}^{t_{1'}}d \bar{1}\Sigma^{<}(1,\bar{1};U)\left[G^{>}(\bar{1},1';U)-G^{<}(\bar{1},1';U)\right].
	\end{multline} 
	\end{widetext}
	We want to use the approximation of the Green's function $G^{\lessgtr}(\mx_1,t_1,\mx_{1'},t_{1'};U)$ for slowly varying applied potential $U(\mathbf{R},T)$ as a function of 
	\begin{equation}
	\mathbf{R}=\frac{\mx_{1}+\mx_{1'}}{2}, \qquad T=\frac{t_1 +t_{1'}}{2}.
	\end{equation}
	We also want to consider $G^{\lessgtr}(\mx_1,t_1,\mx_{1'},t_{1'})$ to be sharply peaked about $\mathbf{r}=0$ and $t=0$, where 
	\begin{equation}
	\mathbf{r}=\mx_{1}-\mx_{1'}, \qquad t= t_1-t_{1'}.
	\end{equation} 
	We can rewrite the Green's function as 
	\begin{equation}
	G^{\lessgtr}(\mx_1,t_1,\mx_{1'},t_{1'};U)=G^{\lessgtr}(\mathbf{r},t,\mathbf{R},T;U).
	\end{equation}
	In the DC case, we only consider the static component of Green's function in which satisfies
	\begin{equation}
	\label{eq:DCHomo}
	\frac{\partial G^{\lessgtr}(\mathbf{r},t,\mathbf{R},T;U)}{\partial T}=0.
	\end{equation}
	We consider the left hand side of of \eqref{eq:Kinetic1}, which we can rewrite as 
	\begin{multline}
	\Big[ i\frac{\p}{\p T} -\mathbf{r}\cdot\nabla_{\mathbf{R}}U(\mathbf{R},T)-t\frac{\p}{\p T} U(\mathbf{R},T)\Big]G^{<}(\mathbf{r},t,\mathbf{R},T;U)\\-G^{<}(\mathbf{r},t,\mathbf{R},T;U)H_{0}(1)+H_0(1')G^{<}(\mathbf{r},t,\mathbf{R},T;U).
	\end{multline}
	We Fourier transform the relative coordinate $\mathbf{r}$ and $t$ by multiplying by $e^{-i\mathbf{p}\cdot\mathbf{r}+i\omega t}$ and integrating over $\mathbf{r}$ and $t$ to obtain 
	\begin{multline}
	\label{eq:Kinetick}
	\!\!\!\!\!\!\!\!\!\!\!\!\!\!\!\!\!\Big[ i\frac{\p}{\p T} -\nabla_{\mathbf{R}}U(\mathbf{R},T)i\nabla_{\mathbf{p}}+\frac{\p}{\p T} U(\mathbf{R},T)i\frac{\p}{\p \omega}\Big]G^{<}(\mathbf{p},\omega,\mathbf{R},T;U)\\-G^{<}(\mathbf{p},\omega,\mathbf{R},T;U)H_{0}(1)+H_0(1')G^{<}(\mathbf{p},\omega,\mathbf{R},T;U).
	\end{multline}
	For an applied static electric field, we have 
	\begin{equation}
	\nabla_{\mathbf{R}}U(\mathbf{R},T)=-e\mathbf{E}, \qquad \frac{\p}{\p T} U(\mathbf{R},T)=0.
	\end{equation}
	Let's look at the approximation that we applied to the Green's function more closely. In the Weyl-Wigner formulation, we only consider the Green's $G^{<}(\mathbf{r},t,\mathbf{R},T;U)$ that is slowly varying in $\mathbf R$. Physically, it means that we only consider the perturbation such that 
	\begin{equation}
	\< c^\dagger_a(\mathbf{R}+\frac{\mathbf r}{2} ) c_b(\mathbf{R}-\frac{\mathbf r}{2} )\>
	\label{eq:c_spatial}
	\end{equation} 	
	is slowly varying in $\mathbf  R$. Fourier transforming \eqref{eq:c_spatial}, we obtain
    \begin{equation}
    \< c^\dagger_a(\mathbf{k}+\frac{\mathbf{K}}{2}) c_b(\mathbf{k}-\frac{\mathbf{K}}{2})\>,
    \end{equation}
	where $\mathbf{k}$ ($\mathbf{K}$) is the momentum conjugate to $\mathbf{r}$ ($\mathbf{R}$). Since we are interested in spatially homogeneous distributions, that means we set $\mathbf{K}=0$ and only consider the Green's function of the form  
	\begin{equation}
	\< c^\dagger_a(\mathbf{k}) c_b(\mathbf{k})\>.
	\end{equation}
	We now can convert the Green's function with sub-lattice indices to the Green's function with band indices using the relation 
	\begin{equation}
	c^f_a(\mathbf{k})=\cU^f_{a\lambda}(\mk) c^f_\lambda(\mathbf{k}), \qquad c^{f\dagger}_a(\mathbf{k})=\cU^{f\dagger}_{\lambda a}(\mk) c^{f\dagger}_\lambda(\mathbf k),
	\end{equation}
	where $\cU^f(\mk)$ and $\cU^{f\dagger}(\mk)$ can be read off from the explicit form of the band wave function in Section \ref{sec:wf} 
	\begin{equation}
	\cU^{f\dagger}(\mk)=\frac{1}{\sqrt{2}}\begin{pmatrix}
	-e^{-2i\theta_{\mk}} & 1\\
	e^{-2i\theta_{\mk}} & 1
	\end{pmatrix}, \cU^{f}(\mk)=\frac{1}{\sqrt{2}}\begin{pmatrix}
	-e^{2i\theta_{\mk}} & e^{2i\theta_{\mk}}\\
	1 & 1
	\end{pmatrix}.
	\end{equation}
	We can transform the equation \eqref{eq:Kinetick} for each flavor
	\begin{widetext}
	\begin{equation}
	\label{eq:KineticBandL1}
	\Big[e\mathbf{E} i\nabla_{\mathbf{p}} \left(\cU(\mathbf{p})g^{<}(\mathbf{p},\omega;U)\cU^\dagger(\mathbf{p})\right) +\frac{1}{m}\cU(\mathbf{p})\begin{pmatrix}
	0 & p^2 g^{<}_{+-}(\mathbf{p},\omega;U)\\
	-p^2 g^{<}_{-+}(\mathbf{p},\omega;U)&0 \end{pmatrix}\cU^\dagger(\mathbf{p})\Big] ,
	\end{equation}
	where we omit $\mathbf{R}$ and $T$ and replace $\partial_{\mathbf{R}}=0$ and $\partial_T=0$, we also denote the Green's function in band indices as 
	\begin{align}
	\label{eq:Unitary1}
	G^{<}(\mathbf{p},\omega;U)= \cU(\mathbf{p}) g^{<}(\mathbf{p},\omega;U) \cU^\dagger(\mathbf{p}),\\
	\label{eq:Unitary2}
	G^{>}(\mathbf{p},\omega;U)= \cU(\mathbf{p}) g^{>}(\mathbf{p},\omega;U) \cU^\dagger(\mathbf{p}).
	\end{align}  
	In the Weyl-Wigner formulation, the LHS of the equation for $G^{>}(\mathbf{p},\omega;U) $ can be written similarly as 
	\begin{equation}
	\label{eq:KineticBandL2}
	\Big[e\mathbf{E} i\nabla_{\mathbf{p}} \left(\cU(\mathbf{p})g^{>}(\mathbf{p},\omega;U)\cU^\dagger(\mathbf{p})\right) +\frac{1}{m}\cU(\mathbf{p})\begin{pmatrix}
	0 & p^2 g^{>}_{+-}(\mathbf{p},\omega;U)\\
	-p^2 g^{>}_{-+}(\mathbf{p},\omega;U)&0 \end{pmatrix}\cU^\dagger(\mathbf{p})\Big] ,
	\end{equation}

	In our calculation, we are interested in the DC transport, in which we omit the off-diagonal part of the density matrix due to the condition \eqref{eq:DCHomo} \footnote{Because of the time-dependence of $c^{(\dagger)}_{\lambda}(\mathbf{k})$, the off-diagonal component of the Green's function $g^{\lessgtr}$ depends on the central time $T$.}. We then linearize the kinetic equation up to linear order in perturbation. The equation \eqref{eq:KineticBandL1} is rewritten as 
	\begin{equation}
	\label{eq:KineticBandL1linear}
	\cU \Big[e\mathbf{E} i\nabla_{\mathbf{p}} g_0^{<}(\mathbf{p},\omega) \Big] \cU^\dagger,
	\end{equation}
	where $g_0^{<}(\mathbf{p},\omega)$ is given by 
	\begin{equation}
	\label{eq:unperurbed_Greens}
	g_0^{<}(\mathbf{p},\omega)=2\pi i\begin{pmatrix}
	\delta(\omega-\epsilon_+(\mathbf{p}))f^0(\epsilon_+(\mathbf{p}),\mu)&0\\
	0&\delta(\omega-\epsilon_-(\mathbf{p}))f^0(\epsilon_-(\mathbf{p}),\mu),
	\end{pmatrix}
	\end{equation}
	where fermionic distribution functions is given by 
	\begin{equation}
	f^0(\epsilon,\mu)=\frac{1}{1+e^{\beta(\epsilon-\mu)}}.
	\end{equation}
	We integrate over $\omega$ to obtain the equal time Green funtion, the equation \eqref{eq:KineticBandL1linear} becomes 
	\begin{equation}
	\label{eq:KineticBandL1linear1}
	\cU \Big[ 2\pi\beta\frac{e\mathbf{E}\cdot\mathbf{p}}{m}\begin{pmatrix}
	f^0(\epsilon_+(\mathbf{p}),\mu)(1-f^0(\epsilon_+(\mathbf{p}),\mu)&0\\
	0&-f^0(\epsilon_-(\mathbf{p}),\mu)(1-f^0(\epsilon_-(\mathbf{p}),\mu))
	\end{pmatrix}\Big] \cU^\dagger,
	\end{equation}
		\end{widetext}
	\subsection{Generalizing the quantum Boltzmann equation to thermal conductivity}
	In order to derive the kinetic equation for thermal conductivity, we turn on a gradient of the temperature $T=T(\mathbf{R})$. The \textit{local} equilibrium distribution function is given by
	\begin{equation}
	f^0_\lambda(\mathbf{p},T(\mathbf{R}),\mu)=\frac{1}{1+e^{\frac{1}{k_BT(\mathbf{R})}\left(\epsilon_\lambda(\mathbf{p}) -\mu\right)}}.
	\end{equation} 
	The equation of motion for $G^{<}(1,1';T(\mathbf{R}))$ is given by \eqref{eq:Kinetic1}.
	We again use Weyl-Wigner formulation and rewrite the left hand side of equation \eqref{eq:Kinetic1} as 
	\begin{widetext}
	\begin{equation}
	\!\!\!\!\!\!\!\label{eq:KineticBandL4}
	\cU \left[ \frac{1}{m}\begin{pmatrix}
	i\left( \mathbf{p}\cdot \nabla_{\mathbf{R}} \right)g^{<}_{++}(\mathbf{p},\omega,\mathbf{R};T(\mathbf{R})) & \left(p^2-\frac{(\nabla_{\mathbf{R}})^2}{4} \right)g^{<}_{+-}(\mathbf{p},\omega,\mathbf{R};T(\mathbf{R}))\\
	-\left(p^2-\frac{(\nabla_{\mathbf{R}})^2}{4} \right) g^{<}_{-+}(\mathbf{p},\omega,\mathbf{R};T(\mathbf{R}))&i\left( -\mathbf{p}\cdot \nabla_{\mathbf{R}} \right)g^{<}_{--}(\mathbf{p},\omega,\mathbf{R};T(\mathbf{R})) \end{pmatrix}\right] \cU^\dagger.
	\end{equation}
	
	If we ignore the off diagonal part of density matrix, equation \eqref{eq:KineticBandL4} can be rewritten as 
	\begin{equation}
	\label{eq:KineticBandL4linear1}
	\cU \left[ \frac{1}{m}\begin{pmatrix}
	i \mathbf{p}\cdot \nabla_{\mathbf{R}} g^{<}_{++}(\mathbf{p},\omega,\mathbf{R};T(\mathbf{R})) & 0\\
	0&-i \mathbf{p}\cdot \nabla_{\mathbf{R}} g^{<}_{--}(\mathbf{p},\omega,\mathbf{R};T(\mathbf{R})) \end{pmatrix}\right] \cU^\dagger.
	\end{equation}
	Up to linear order in perturbation, we replace $g^{<}(\mathbf{p},\omega,\mathbf{R};T(\mathbf{R}))$ by the equilibrium one
	\begin{equation}
	g^{<}_{0\lambda\lambda}(\mathbf{p},\omega,\mathbf{R};T(\mathbf{R}))=2\pi i\delta(\omega-\epsilon_\lambda(\mathbf{p}))f^0_\lambda(\mathbf{p},T(\mathbf{R}),\mu).
	\end{equation}
	Again, we integrate over $\omega$ to obtain the equal time Green's function. We consider a constant gradient in temperature by introducing the space-time independent driving \textit{force}
	$\mathbf{F}^T=-\frac{\nabla_{\mathbf{R}}T}{T}$, the equation \eqref{eq:KineticBandL4linear1} becomes 
	\small
	\begin{equation}
	\!\!\!\!\!\!\!\!\!\!\!\!\!\!\!\!\label{eq:KineticBandL4linear3}
	\cU \left[ \frac{2\pi\beta}{m}\begin{pmatrix}
	\mathbf{p}\cdot \mathbf{F}^T(\epsilon_+(\mathbf{p})-\mu) f_0(\epsilon_+(\mathbf{p}),\mu)(1-f_0(\epsilon_+(\mathbf{p}),\mu)) & 0\\
	0&-\mathbf{p}\cdot \mathbf{F}^T(\epsilon_-(\mathbf{p})-\mu) f_0(\epsilon_-(\mathbf{p}),\mu)(1-f_0(\epsilon_-(\mathbf{p}),\mu)) \end{pmatrix}\right] \cU^\dagger.
	\end{equation}
	\normalsize
	\end{widetext}
	\subsection{Generalizing the quantum Boltzmann equation to shear viscosity}
	In order to derive the kinetic equation for shear viscosity, we assume that the particles and holes have a spatially dependent local velocity. We consider the \textit{local} equilibirium distribution function 
	\begin{equation}
	f^0_\lambda(\mathbf{p},\mathbf{u}_\lambda(\mathbf{R}),\mu)=\frac{1}{1+e^{\beta(\epsilon_\lambda(\mathbf{p})-\mathbf{u}_\lambda(\mathbf R)\cdot\mathbf{p}-\mu)}}.
	\end{equation}
	We can follow the last subsection to obtain the kinetic equation for $G^{<}(1,1';\{\mathbf{u}_\lambda\})$ \eqref{eq:Kinetic1}. We can use Weyl-Wigner coordinate and obtain the left hand side of equation \eqref{eq:Kinetic1} as 
	\begin{widetext}
	\begin{equation}
	\label{eq:KineticBandL3}
	\cU \left[ \begin{pmatrix}
	i\left( \frac{\mathbf{p}}{m}\cdot \nabla_{\mathbf{R}} \right)g^{<}_{++}(\mathbf{p},\omega,\mathbf{R};\{\mathbf{u}_\lambda\}) & \left(\frac{p^2}{m}-\frac{(\nabla_{\mathbf{R}})^2}{4m} \right) g^{<}_{+-}(\mathbf{p},\omega,\mathbf{R};\{\mathbf{u}_\lambda\})\\
	\left(-\frac{p^2}{m}+\frac{(\nabla_{\mathbf{R}})^2}{4m} \right) g^{<}_{-+}(\mathbf{p},\omega,\mathbf{R};\{\mathbf{u}_\lambda\})&i\left( -\frac{\mathbf{p}}{m}\cdot \nabla_{\mathbf{R}} \right)g^{<}_{--}(\mathbf{p},\omega,\mathbf{R};\{\mathbf{u}_\lambda\}) \end{pmatrix}\right] \cU^\dagger.
	\end{equation}
	If we ignore the off-diagonal part of the density matrix and consider the linearized version of equation \eqref{eq:KineticBandL3}, we obtain 
	\begin{equation}
	\label{eq:KineticBandL3Linear}
	\cU \left[ \frac{1}{m}\begin{pmatrix}
	i \mathbf{p}\cdot \nabla_{\mathbf{R}} g^{<}_{++}(\mathbf{p},\omega,\mathbf{R};\{\mathbf{u}_\lambda\}) & 0\\
	0&-i \mathbf{p}\cdot \nabla_{\mathbf{R}}g^{<}_{--}(\mathbf{p},\omega,\mathbf{R};\{\mathbf{u}_\lambda\}) \end{pmatrix}\right] \cU^\dagger.
	\end{equation}
	Up to linear order in perturbation, we replace $g^{<}_{\lambda\lambda}(\mathbf{p},\omega,\mathbf{R};\{\mathbf{u}_\lambda\})$ by the equilibrium one 
	\begin{equation}
	g^{<}_{0\lambda\lambda}(\mathbf{p},\omega,\mathbf{R};\{\mathbf{u}_\lambda\})=2\pi i\delta(\omega-\epsilon_\lambda(\mathbf{p}))f^0_\lambda(\mathbf{p},\mathbf{u}_\lambda(\mathbf{R}),\mu).
	\end{equation} 
	Again, we integrate over $\omega$ to obtain equal time Green's function, the equation \eqref{eq:KineticBandL3Linear} becomes
	\begin{equation}
	\label{eq:KineticBandL3Linear1}
	\cU \left[ \frac{2\pi\beta}{m}\begin{pmatrix}
	-p^ip^j \partial_i u_{+j} f^0(\epsilon_+(\mathbf{p}),\mu)(1-f^0(\epsilon_+(\mathbf{p}),\mu)) & 0\\
	0& p^ip^j \partial_i u_{-j}f^0(\epsilon_-(\mathbf{p}),\mu)(1-f^0(\epsilon_-(\mathbf{p}),\mu)) \end{pmatrix}\right] \cU^\dagger.
	\end{equation}
	\end{widetext}
	We turn on the shear which, by definition, is the divergence-free background flow $\partial_i u_{\lambda}^i=0$.
	We define the shear tensor 
	\begin{equation}
	X^\lambda_{ ij}=\frac{1}{2}\left( \partial_i u_{\lambda}^j+\partial_j u_{\lambda}^i\right).
	\end{equation}
	We can turn on the space-time independent off-diagonal part of the shear tensor so that
	\begin{equation}
	X^\lambda_{ 12}=X^\lambda_{ 21}=F_\lambda, \qquad X^\lambda_{ 11}=X^\lambda_{ 22}=0,
	\end{equation} 
	where $F_\lambda$ is space-time independent. The equation \eqref{eq:KineticBandL3Linear} becomes
	\begin{widetext}
	\begin{equation}
	\label{eq:KineticBandL3Linear2}
	\cU \left[ \frac{2\pi\beta}{m}\begin{pmatrix}
	-2p^1p^2 F_+ f^0(\epsilon_+(\mathbf{p}),\mu)(1-f^0(\epsilon_+(\mathbf{p}),\mu)) & 0\\
	0& 2p^1p^2 F_-f^0(\epsilon_-(\mathbf{p}),\mu)(1-f^0(\epsilon_-(\mathbf{p}),\mu)) \end{pmatrix}\right] \cU^\dagger.
	\end{equation}
	\end{widetext}
	We need to solve the kinetic equations and calculate equal time Green's function  $g^{<}_\lambda(\mathbf{p};\{F_\lambda\})$, and derive the stress tensor $T_\lambda^{ij}$. The viscosity can be read off from the equation 
	\begin{equation}
	T_\lambda^{12}=-\eta_{\lambda \lambda'}F_{\lambda'}.
	\end{equation}
	

	
	\subsection{The collision integral induced by Coulomb interaction}
	In this section, we will derive the right-hand side of \eqref{eq:Kinetic1} in the Weyl-Wigner formulation. Let's simplify the collisional part of the kinetic equation. With the assumption that, in Weyl-Wigner coordinates, $G^{\lessgtr}(\mathbf{r},t,\mathbf{R},T)$ and $\Sigma^{\lessgtr}(\mathbf{r},t,\mathbf{R},T)$ vary slowly in $\mathbf{R}$ and $T$, we can perform the Fourier transformation and rewrite the collision integral. In the rest of this section, we will omit $X$,$\mathbf{R}$ and $T$ to simplify the notation. We define the following notation
	\begin{widetext}
	\begin{equation}
	\Sigma^{>,<}(\mathbf{p},\omega,\mathbf{R},T;\{X\})=\int d\mathbf{r} dt e^{-i\mathbf{p}\mathbf{r}+i\omega t}\Sigma^{>,<}(\mathbf{r},t,\mathbf{R},T;\{X\}).
	\end{equation}
	
	The explicit formula of self energy is given by 
	\begin{multline}
	\Sigma^{>,<}_{\alpha\beta}(\mk,\omega)=(2\pi)^3\int \frac{d^2\mk_1}{(2\pi)^2}\frac{d\omega_1}{2\pi}\frac{d^2\mk_2}{(2\pi)^2}\frac{d\omega_2}{2\pi}\frac{d^2\mk_3}{(2\pi)^2}\frac{d\omega_3}{2\pi}\delta(\mk+\mk_1-\mk_2-\mk_3)\delta(\omega+\omega_1-\omega_2-\omega_3)\\
	\times \Big[V_C(\mk-\mk_2)V_C(\mk-\mk_2)[G^{>,<}_{\alpha\beta}(\mk_2,\omega_2)G^{>,<}_{\gamma\delta}(\mk_3,\omega_3)G^{<,>}_{\delta\gamma}(\mk_1,\omega_1)\\
	-V_C(\mk-\mk_2)V_C(\mk-\mk_3)G^{>,<}_{\alpha\gamma}(\mk_2,\omega_2)G^{<,>}_{\gamma\delta}(\mk_1,\omega_1)G^{>,<}_{\delta\beta}(\mk_3,\omega_3)\Big].
	\end{multline}
	
	Since flavor index is conserved at each vertex, in the following formulae, we will omit the flavor index. The transformation of the Green's function between sub-lattice index and band index is given by \eqref{eq:Unitary1} and \eqref{eq:Unitary2}. We can rewrite the self energy as 
	\begin{multline}
	\Sigma^{>,<}_{\alpha\beta}(\mk,\omega)=(2\pi)^3\int \frac{d^2\mk_1}{(2\pi)^2}\frac{d\omega_1}{2\pi}\frac{d^2\mk_2}{(2\pi)^2}\frac{d\omega_2}{2\pi}\frac{d^2\mk_3}{(2\pi)^2}\frac{d\omega_3}{2\pi}\delta(\mk+\mk_1-\mk_2-\mk_3)\delta(\omega+\omega_1-\omega_2-\omega_3)\\
	\times \Big[N_fV_C(\mk-\mk_2)V_C(\mk-\mk_2)[\cU_{\mk_2}g^{>,<}(\mk_2,\omega_2)\cU^\dagger_{\mk_2}]_{\alpha\beta}[\cU_{\mk_3}g^{>,<}(\mk_3,\omega_3)\cU^\dagger_{\mk_3}]_{\gamma\delta}[\cU_{\mk_1}g^{<,>}(\mk_1,\omega_1)\cU^\dagger_{\mk_1}]_{\delta\gamma}\\
	-V_C(\mk-\mk_2)V_C(\mk-\mk_3)[\cU_{\mk_2}g^{>,<}(\mk_2,\omega_2)\cU^\dagger_{\mk_2}]_{\alpha\gamma}[\cU_{\mk_1}g^{<,>}(\mk_1,\omega_1)\cU^\dagger_{\mk_1}]_{\gamma\delta}[\cU_{\mk_3}g^{>,<}(\mk_3,\omega_3)\cU^\dagger_{\mk_3}]_{\delta\beta}\Big],
	\end{multline}
	\end{widetext}
	where $\alpha$,$\beta$,$\gamma$ and $\delta$ now are sub-lattice indices only. The factor of $N_f$ in the first term comes from the summation over the flavor index of the loop diagram. The simplification comes from the assumption that the perturbed Green's functions for each flavor are the same. 
	Now we consider only the diagonal part of the density matrix in band index
	\begin{align}
	g^<_{\lambda\lambda'}(\mk,\omega)=i2\pi\delta(\omega-\epsilon_\lambda(\mk))f_\lambda(\mk)\delta_{\lambda \lambda'},\\
	g^>_{\lambda\lambda'}(\mk,\omega)=-i2\pi\delta(\omega-\epsilon_\lambda(\mk))(1-f_\lambda(\mk))\delta_{\lambda \lambda'}.
	\end{align}
	The right hand side of \eqref{eq:Kinetic1} for each band index $\lambda$, after integrating over $\omega$ and multiplying by $\cU^\dagger(\mathbf{p})$ on the left and $\cU(\mathbf{p})$ on the right, is the collision integral for this band index and given by
	\begin{equation}
	\label{eq:CoIA}
	2\pi I_\lambda[\{f_{\lambda_i}(\mk_i)\}](\mathbf{p})=-2\pi(-if_\lambda(\mathbf{p})\sigma^>_{\lambda\lambda}(\mathbf{p})-i(1-f_\lambda(\mathbf{p}))\sigma^<_{\lambda\lambda}(\mathbf{p})),
	\end{equation} 
	where
	\begin{widetext}
	\begin{multline}
	\label{eq:selfEn1}
	\sigma^>_{\lambda\lambda}(\mathbf{p})=-i(2\pi)^3\sum_{\lambda_1\lambda_2\lambda_3}\int \frac{d^2\mk_1}{(2\pi)^2}\frac{d^2\mk_2}{(2\pi)^2}\frac{d^2\mk_3}{(2\pi)^2}\delta(\mathbf{p}+\mk_1-\mk_2-\mk_3)\delta(\epsilon_\lambda(\mathbf{p})+\epsilon_{\lambda_1}(\mk_1)-\epsilon_{\lambda_2}(\mk_2)-\epsilon_{\lambda_3}(\mk_3))\\
	\times \Big[N_f V_C(\mathbf{p}-\mk_2)V_C(\mathbf{p}-\mk_2) M_{\lambda_3\lambda_1}(\mk_3,\mk_1)M_{\lambda_1\lambda_3}(\mk_1,\mk_3)M_{\lambda\lambda_2}(\mathbf{p},\mk_2)M_{\lambda_2\lambda}(\mk_2,\mathbf{p})f_{\lambda_1}(\mk_1)(1-f_{\lambda_2}(\mk_2))(1-f_{\lambda_3}(\mk_3))\\
	-V_C(\mathbf{p}-\mk_2)V_C(\mathbf{p}-\mk_3) M_{\lambda_1\lambda_3}(\mk_1,\mk_3)M_{\lambda_2\lambda_1}(\mk_2,\mk_1)M_{\lambda\lambda_2}(\mathbf{p},\mk_2)M_{\lambda_3\lambda}(\mk_3,\mathbf{p})f_{\lambda_1}(\mk_1)(1-f_{\lambda_2}(\mk_2))(1-f_{\lambda_3}(\mk_3))\Big],
	\end{multline}
	and
	\begin{multline}
	\label{eq:selfEn2}
	\sigma^<_{\lambda\lambda}(\mathbf{p})=i(2\pi)^3\sum_{\lambda_1\lambda_2\lambda_3}\int \frac{d^2\mk_1}{(2\pi)^2}\frac{d^2\mk_2}{(2\pi)^2}\frac{d^2\mk_3}{(2\pi)^2}\delta(\mathbf{p}+\mk_1-\mk_2-\mk_3)\delta(\epsilon_\lambda(\mathbf{p})+\epsilon_{\lambda_1}(\mk_1)-\epsilon_{\lambda_2}(\mk_2)-\epsilon_{\lambda_3}(\mk_3))\\
	\times \Big[N_f V_C(\mathbf{p}-\mk_2)V_C(\mathbf{p}-\mk_2) M_{\lambda_3\lambda_1}(\mk_3,\mk_1)M_{\lambda_1\lambda_3}(\mk_1,\mk_3)M_{\lambda\lambda_2}(\mathbf{p},\mk_2)M_{\lambda_2\lambda}(\mk_2,\mathbf{p})(1-f_{\lambda_1}(\mk_1))f_{\lambda_2}(\mk_2)f_{\lambda_3}(\mk_3)\\
	-V_C(\mathbf{p}-\mk_2)V_C(\mathbf{p}-\mk_3) M_{\lambda_1\lambda_3}(\mk_1,\mk_3)M_{\lambda_2\lambda_1}(\mk_2,\mk_1)M_{\lambda\lambda_2}(\mathbf{p},\mk_2)M_{\lambda_3\lambda}(\mk_3,\mathbf{p})(1-f_{\lambda_1}(\mk_1))f_{\lambda_2}(\mk_2)f_{\lambda_3}(\mk_3)\Big].
	\end{multline}	 
	\end{widetext}
	Combining equations \eqref{eq:CoIA}, \eqref{eq:selfEn1} and \eqref{eq:selfEn2} gives us the contribution to collision integral from the Coulomb interaction between quasi-particles \eqref{eq:CoI}.

	\section{Detailed derivation of \eqref{eq:CoD}}
	\label{sec:CoD}
	Fourier transforming \eqref{eq:H_dis} and writing this in terms of the creation and annihilation operators, we find
	\begin{multline}
	\label{eq:HamD}
	H_{\textrm{dis}}=\frac{1}{L^2}\sum_i\sum_f\sum_{\lambda_1,\lambda_2}\sum_{\vec{k}_1}\sum_{\vec{k}_2}\tilde V_{\lambda_1,\lambda_2}(\vec{k_1},\vec{k_2})\\\times e^{i\vec{x_i}\cdot(\vec{k_1}-\vec{k_2})}\hat c_{\lambda_1f}^\dagger(\vec{k_1})\hat c_{\lambda_2f}(\vec{k_2}),
	\end{multline}
	where
	\begin{equation}
	\tilde V_{\lambda_1,\lambda_2}(\vec{k_1},\vec{k_2})=V_C(\vec{k_1}-\vec{k_2})\mathcal{M}_{\lambda_1,\lambda_2}(\vec{k_1},\vec{k_2}).
	\end{equation}
	From the interacting Hamiltonian \eqref{eq:HamD}, we work out the matrix element 
	\begin{align}
	&\Braket{\vec{k}\lambda f|\hat H_{\textrm{dis}}|\vec{k}'\lambda f}\nonumber\\&=\Braket{FS|\hat c_{\lambda f}(\vec{k})\hat H_{\textrm{dis}}\hat c^\dagger_{\lambda f}(\vec{k}')|FS}\\&=\frac{1}{L^2}\tilde V_{\lambda,\lambda}(\vec{k},\vec{k}')\sum_i e^{i\vec{x_i}\cdot(\vec{k}-\vec{k}')}.\nonumber
	\end{align}
	We square the matrix element and average over disorder realizations with $N_\textrm{imp}$ impurities
	\begin{align}
	&\bigg|\Braket{\vec{k}\lambda f|\hat H_{\textrm{dis}}|\vec{k}'\lambda f}\bigg|^2\nonumber\\&=\frac{1}{L^4}N_\textrm{imp}|\tilde V_{\lambda,\lambda}(\vec{k},\vec{k}')|^2\\&=\frac{1}{L^2}n_\textrm{imp}|\tilde V_{\lambda,\lambda}(\vec{k},\vec{k}')|^2.\nonumber
	\end{align}
	From Fermi's Golden Rule, the scattering rate is
	\begin{equation}
	\Gamma(\vec{k}\rightarrow\vec{p})=2\pi \bigg|\Braket{\vec{k}\lambda f|\hat H_{\textrm{dis}}|\vec{p}\lambda f}\bigg|^2\times\frac{dN}{dE},
	\end{equation}
	where the density of states in 2d is 
	\begin{equation}
	\frac{dN}{dE}=\frac{mL^2}{2\pi}, 
	\end{equation}
	so
	\begin{equation}
	\Gamma(\vec{k}\rightarrow\vec{p})=2\pi n_\textrm{imp}|\tilde V_{\lambda,\lambda}(\vec{k},\vec{p})|^2\times\frac{m}{2\pi}.
	\end{equation}
	The collision integral for the impurity scattering is
	\begin{multline}
	I_{\textrm{dis}}(\vec{p})=\frac{2\pi}{m}\int\frac{d^2k}{(2\pi)^2}\delta(\epsilon_\lambda(p)-\epsilon_\lambda(k))\\\times\bigg(\Gamma(\vec{k}\rightarrow\vec{p})f(\vec{k})(1-f(\vec{p}))-\Gamma(\vec{p}\rightarrow\vec{k})f(\vec{p})(1-f(\vec{k}))\bigg),
	\end{multline}
	\begin{multline}
	I_{\textrm{dis}}(\vec{p})=2\pi n_\textrm{imp}\int\frac{d^2\vec{k}}{(2\pi)^2}\delta(\epsilon_\lambda(p)-\epsilon_\lambda(k))|\tilde V_{\lambda\lambda}(\vec{p},\vec{k})|^2\\\times\bigg(f_\lambda(\vec{p})(1-f_\lambda(\vec{k}))-f_\lambda(\vec{k})(1-f_\lambda(\vec{p}))\bigg).
	\end{multline}
	Now write 
	\begin{equation}
	f_\lambda(\vec{p})=f_\lambda^0(p)+f_\lambda^0(p)[1-f_\lambda^0(p)]h_\lambda(\vec{p})
	\end{equation}
	and linearize the collision integral (in this case the exact collision integral is already linear in $h$).
	\begin{multline}
	I^{(1)}_{\textrm{dis}}[h_{\lambda_i} (\vec{k}_i)](\vec{p})=2\pi n_{\textrm{imp}}f_\lambda^0(p)[1-f_\lambda^0(p)]\\\times\int\frac{d^2\vec{k}}{(2\pi)^2}\delta(\epsilon_\lambda(p)-\epsilon_\lambda(k))|\tilde V_{\lambda\lambda}(\vec{p},\vec{k})|^2\bigg(h_\lambda(\vec{p})-h_\lambda(\vec{k})\bigg).
	\end{multline}
	Now let us take the fully screened potential
	\begin{equation}
	\tilde V_{\lambda_1,\lambda_2}(\vec{k_1},\vec{k_2})=\frac{2\pi Ze^2}{\epsilon_rq_{TF}}\mathcal{M}_{\lambda_1,\lambda_2}(\vec{k_1},\vec{k_2})
	\end{equation}
	to obtain
	\begin{multline}
	I^{(1)}_{\textrm{dis}}[h_{\lambda_i} (\vec{k}_i)](\vec{p})=2\pi n_{\textrm{imp}}\bigg(\frac{2\pi Ze^2}{\epsilon_rq_{TF}}\bigg)^2f_\lambda^0(p)[1-f_\lambda^0(p)]\\\times\int\frac{d^2\vec{k}}{(2\pi)^2}\delta(\epsilon_\lambda(p)-\epsilon_\lambda(k))|\mathcal{M}_{\lambda\lambda}(\vec{p},\vec{k})|^2\bigg(h_\lambda(\vec{p})-h_\lambda(\vec{k})\bigg).
	\end{multline}
	Using 
	\begin{equation}
	\delta(\epsilon_\lambda(p)-\epsilon_\lambda(k))=\delta\bigg(\frac{p^2}{2m}-\frac{k^2}{2m}\bigg)=\frac{m}{k}\delta(p-k)
	\end{equation}
	leads to
	\begin{multline}
	I^{(1)}_{\textrm{dis}}[h_{\lambda_i} (\vec{k}_i)](\vec{p})=\alpha f_\lambda^0(p)[1-f_\lambda^0(p)]\int\frac{d\theta_k}{2\pi}|\mathcal{M}_{\lambda\lambda}(\vec{p},p\hat{\vec{k}})|^2\\\times\bigg(h_\lambda(\vec{p})-h_\lambda(p\hat{\vec{k}})\bigg),
	\end{multline}
	where we have defined
	\begin{equation}
	\kappa=mn_{\textrm{imp}}\bigg(\frac{2\pi Ze^2}{\epsilon_rq_{TF}}\bigg)^2
	\end{equation}
	and
	\begin{multline}
	I^{(1)}_{\textrm{dis}}[h_{\lambda_i} (\vec{k}_i)](\vec{p})=\kappa f_\lambda^0(p)[1-f_\lambda^0(p)]\int\frac{d\theta_k}{2\pi}\cos^2(\theta_k-\theta_p)\\\times\bigg(h_\lambda(\vec{p})-h_\lambda(p\hat{\vec{k}})\bigg).
	\end{multline}
	\begin{equation}
	h_\lambda(\mathbf{p})=\beta\frac{e\mathbf{E}}{m}\cdot\bigg(\mathbf{p}\chi_\lambda^\parallel(p)+\mathbf{p}\times\hat{\vec{z}}\chi_\lambda^\perp(p)\bigg),
	\end{equation}
	which gives us equation \eqref{eq:CoD}.

	\section{Detailed numerical calculations}
In this section, we explain in detail our numerical setup as well as the steps of calculations. In particular, we explain how to turn the QBE into a matrix equation. We follow \cite{Mueller2008}. The Boltzmann equation is
	\begin{multline}
	\label{eq:BolE2}
	-\lambda \beta\frac{e\mathbf{E}\cdot\mathbf{p}}{m} f^0_\lambda(\mathbf{p})[1-f^0_\lambda(\mathbf{p})]\\+\frac{eB\lambda}{m}f^0_\lambda(\mathbf{p})[1-f^0_\lambda(\mathbf{p})](\vec p\times \vec{\hat{z}})\cdot \partial_{\vec p}h_\lambda(\mathbf{p})\\=-I_\lambda[\{h_{\lambda_i}(\vec k_i)\}](\mathbf{p})
	\end{multline}
The suggested ansatz in this calculation is 
	\begin{equation}
	\label{eq:as1}
	h_\lambda(\mathbf{p})=\beta\frac{e\mathbf{E}}{m}\cdot\bigg(\mathbf{p}\chi_\lambda^\parallel(p)+\mathbf{p}\times\hat{\vec{z}}\chi_\lambda^\perp(p)\bigg)
	\end{equation}
Expand in terms of basis functions
	\begin{equation}	
		\chi_\lambda^{\parallel,\perp}(k)=\beta\sum_n a_n^{\parallel,\perp} g_n(\lambda,k)
	\end{equation}
such that $a$ is dimensionless. Here the basis functions are taken to be
	\begin{multline}
		g_n(\lambda,k)=1,\lambda,K,\lambda K,K^2,\lambda K^2,\\K^3e^{-K/2},\lambda K^3e^{-K/2}...K^Ne^{-K/2},\lambda K^Ne^{-K/2}
	\end{multline}
where $K=\sqrt{\beta/m}k$ is the dimensionless momentum. For all powers $n>2$ we multiply by an exponential factor so the basis function is $K^ne^{-K/2}$. We expand in up to 16 basis functions. Increasing the number of basis function changes the results only marginally. 
Use the fact that this must be valid for all $\vec E$, sum over $\lambda$, multiply separately by $\vec{\hat{p}}g_m(\lambda,p)$ and $(\vec{\hat{p}}\times \vec{\hat{z}})g_m(\lambda,p)$ and integrate over $\vec p$. This yields two equations that can be summarized in matrix form as
\begin{equation}
	\begin{pmatrix}
	M & -B\\
	B& M
	\end{pmatrix}
	\begin{pmatrix}
	\vec a^\parallel\\
	\vec a^\perp
	\end{pmatrix}=
	\begin{pmatrix}
	\vec F\\
	\vec 0
	\end{pmatrix}
	\label{eq:matrixEq}
\end{equation}
	\begin{widetext}
where we defined the dimensionless matrices
\begin{equation}
	M_{mn}=\beta \bigg(\frac{\beta}{m}\bigg)^{3/2}\sum_\lambda\int \frac{d^2\mathbf{p}}{(2\pi)^2}g_m(\lambda,p) I_\lambda\bigg[\bigg\{\vec{\hat{p}}\cdot\mathbf{k}_ig_n(\lambda_i,k_i)\bigg\}\bigg](\mathbf{p})
\end{equation}
and
\begin{equation}
	B_{mn}=\beta \bigg(\frac{\beta}{m}\bigg)^{3/2}\sum_\lambda\int \frac{d^2\mathbf{p}}{(2\pi)^2}\frac{eB\lambda}{m}f^0_\lambda(p)[1-f^0_\lambda(p)]p  g_n(\lambda,p)g_m(\lambda,p)
\end{equation}
and the dimensionless vector
\begin{equation}
	F_m= \bigg(\frac{\beta}{m}\bigg)^{3/2}\sum_\lambda\int \frac{d^2\mathbf{p}}{(2\pi)^2} \lambda p f^0_\lambda(p)[1-f^0_\lambda(p)]g_m(\lambda,p)
\end{equation}
\eqref{eq:matrixEq} can be inverted to yield
\begin{equation}
\begin{pmatrix}
\vec a^\parallel\\
\vec a^\perp
\end{pmatrix}=
\begin{pmatrix}
K & \bar K\\
-\bar K& K
\end{pmatrix}
\begin{pmatrix}
\vec F\\
\vec 0
\end{pmatrix}
\end{equation}
where
\begin{equation}
	K=(M+BM^{-1}B)^{-1}, \qquad
	\bar K=M^{-1}B(M+BM^{-1}B)^{-1}.
\end{equation}
The charge current is
	
	\begin{align}
	\mathbf{J}&=\frac{e}{m}N_f\sum_\lambda \int \frac{d^2\mathbf{p}}{(2\pi)^2}\lambda \mathbf{p}f_\lambda(\mathbf{p})\\&=\beta N_f\sum_\lambda \int \frac{d^2\mathbf{p}}{(2\pi)^2}\lambda \mathbf{p}f^0_\lambda(\mathbf{p})[1-f^0_\lambda(\mathbf{p})]\frac{e^2\mathbf{E}}{m^{*2}}\nonumber\\
	&\cdot \bigg(\mathbf{p}\chi_\lambda^\parallel(p)+\mathbf{p}\times\hat{\vec{z}}\chi_\lambda^\perp(p)\bigg).\nonumber
	\end{align}
	The DC conductivity is read off as 
	\begin{equation}
	\sigma_{xx}=\beta N_f\sum_\lambda \int \frac{d^2\mathbf{p}}{(2\pi)^2}\lambda f^0_\lambda(p)[1-f^0_\lambda(p)]\frac{e^2p_x^2}{m^{*2}}\chi_\lambda^\parallel(p)=\frac{N_f e^2}{2\hbar}\vec G\cdot K \vec F,
	\end{equation}
	\begin{equation}
	\sigma_{xy}=-\beta N_f\sum_\lambda \int \frac{d^2\mathbf{p}}{(2\pi)^2}\lambda f^0_\lambda(p)[1-f^0_\lambda(p)]\frac{e^2p_x^2}{m^{*2}}\chi_\lambda^\perp(p)=\frac{N_f e^2}{2\hbar}\vec G\cdot\bar K \vec F.
	\end{equation}
where we have exceptionally restored $\hbar$ and where the dimensionless vector
\begin{equation}
G_m=\bigg(\frac{\beta}{m}\bigg)^{2}\sum_\lambda\int \frac{d^2\mathbf{p}}{(2\pi)^2} \lambda p^2 f^0_\lambda(p)[1-f^0_\lambda(p)]g_m(\lambda,p)
\end{equation}

		\end{widetext}
\bibliography{BiGraphene}

\end{document}